\documentclass[12pt]{iopart}

\newcommand{\mez}{\hspace*{+0.50cm}}
\newcommand{\mz}{\hspace*{+0.25cm}}
\newcommand{\m}{\hspace*{-0.50mm}}
\newcommand{\n}{\hspace*{-0.25mm}}
\newcommand{\be}{\begin{equation}}
\newcommand{\ee}{\end{equation}}
\newcommand{\la}{\langle}
\newcommand{\ra}{\rangle}
\usepackage{iopams,cite}
\usepackage{bm,amssymb}
\usepackage{graphicx,color}
\usepackage[symbol]{footmisc}

\begin{document}

\title[Equations of motion for the exceptional points]{Equations of motion governing the dynamics of the exceptional points
of parameterically dependent nonhermitian Hamiltonians}

\author{Milan \v{S}indelka}
\address{Institute of Plasma Physics of the Czech Academy of Sciences, Za Slovankou 1782/3, 18200 Prague, Czech Republic}
\ead{sindelka@ipp.cas.cz}

\author{Pavel Str\'{a}nsk\'{y} and Pavel Cejnar}
\address{Institute of Nuclear and Particle Physics, Faculty of Mathematics and Physics, Charles University, V Hole\v{s}ovi\v{c}k\'{a}ch 2, 18000 Prague, Czech Republic}
\ead{stransky@ipnp.troja.mff.cuni.cz, cejnar@ipnp.mff.cuni.cz}

\vspace{10pt}
\begin{indented}
\item[]\today
\end{indented}

\begin{abstract}
We study exceptional points (EPs) of a nonhermitian Hamiltonian $\hat{H}\n(\lambda,\delta)$ whose parameters $\lambda \in {\mathbb C}$
and $\delta \in {\mathbb R}$. As the real control parameter $\delta$ is varied, the $k$-th EP (or $k$-th cluster of simultaneously existing EPs)
of $\hat{H}\n(\lambda,\delta)$ moves in the complex plane of $\lambda$ along a continuous trajectory, $\lambda_k(\delta)$. We derive a self
contained set of equations of motion (EOM) for the trajectory $\lambda_k(\delta)$, while interpreting $\delta$ as the propagation time.
Such EOM become of interest whenever one wishes to study the response of EPs to external perturbations or continuous parametric changes of the pertinent Hamiltonian. This is e.g.~the case of EPs emanating from hermitian curve crossings/degeneracies (which turn into avoided
crossings/near-degeneracies when the Hamiltonian parameters are continuously varied). The presented EOM for EPs have not only their theoretical
merits, they possess also a substantial practical relevance. Namely, the just presented approach can be regarded even as an efficient numerical
method, useful for generating EPs for a broad class of complex quantum systems encountered in atomic, nuclear and condensed matter physics. Performance of such a method is tested here numerically on a simple yet nontrivial toy model.
\end{abstract}

\vspace{2pc}

\submitto{\JPA}

\maketitle

\section{Introduction}

Nonhermitian Hamiltonians give rise to a special kind of degeneracies (the so called exceptional points, EPs) which are not encountered within
the standard hermitian quantum mechanics. Namely, not only the (complex) eigenvalues, but also the corresponding eigenvectors become degenerate
(coalescent) at the EP \cite{Kato,Bender,Moiseyev,NM-Book,Mailybaev-1,Miri}. Mathematical peculiarities of such a situation include self-orthogonality, an unusual closure property,
and multivaluedness of the involved eigenvalues when encircling an EP in the parameter space of the Hamiltonian. Importantly, the EPs arise not
only in toy models, but also in a vast amount of physically relevant and experimentally accessible contexts (quantum mechanics of laser driven
atoms, waveguide optics, acoustics, electric circuit theory, elasticity)
%, single mode lasing, unidirectional transmission and reflection, enhanced sensing)
where they imply surprising counter-intuitive phenomena, see
e.g.~short reviews \cite{Heiss-6,Rotter,Heiss-7} and also Refs.~\cite{Zirnbauer,Heiss-1,Heiss-2,Berry,Garmon,Klaiman,EPs-examples-1,EPs-examples-2,EPs-examples-3,EPs-examples-4,EPs-examples-5,EPs-Milan-Summerschool,EPs-Milan-WGs,EPs-Petra-Milan,EPs-Petra-alone,single-mode-lasing-1,single-mode-lasing-2,utor-1,utor-2,utor-3,es-1,es-2,es-3,es-4}. Moreover, relevance of EPs in the context of quantum chaos and quantum phase transitions has been demonstrated theoretically \cite{Heiss-1,Heiss-3,Heiss-4,Heiss-5,Cejnar-1,Cejnar-2,Cejnar-2b,Lee,Borisov,Znojil-1,Wang,Lea}.
The role of EPs in the superradiance phenomenon has also been recognized \cite{Jung,Cejnar-3}. Higher order EPs have been explored e.g.~in
Refs.~\cite{Graefe,Demange,Znojil-2}.

Substantial effort has been invested into developing computational methods for finding the EPs explicitly for a given nonhermitian Hamiltonian
\cite{Mailybaev-2,Uzdin,Kirillov,Nennig}. In spite of great ingenuity and insighfulness of such algorithms, their application to concrete systems is not
always straightforward. Difficulties arise especially when the Hamiltonian under study supports and existence of many EPs, which are typically
associated with avoided crossings encountered in the framework of the pertinent hermitian theory.

The purpose of our present article is to further contribute both to the theory and computational methodology related to EPs. Namely, we study
the response of EPs to continuous changes of the Hamiltonian parameters. Our intention is then to calculate EPs of a given problem by means of
a continuous parametric propagation, starting from an arrangement where the EPs are trivial (or at least easy) to find.

Our basic idea can be sketched as follows.
Let us consider a parameterically $\lambda$-dependent Hamiltonian of the general form
\be \label{hat-H-lambda-take-1}
   \hat{H}(\lambda) \; = \; \hat{H}_0 \; + \; \lambda \, \hat{V} \mez . \mez (\lambda \in {\mathbb C})
\ee
We are looking for the EPs of $\hat{H}(\lambda)$ in the complex $\lambda$-plane. The present paper pursues the following strategy:
We conveniently express $\hat{V}$ as a sum $\hat{V}=\hat{V}_0+\hat{V}_1$, where the component $\hat{V}_0$ is chosen so
that the eigenvalue problem and the EPs of $\hat{H}_0+\lambda\hat{V}_0$ are either trivially resolvable or at least easy to handle.
A prototypical example (which will be elaborated fully explicitly below in Section 3) corresponds to cases when $[\hat{H}_0,\hat{V}_0]=\hat{0}$.
Then $\hat{H}_0$ and $\hat{V}_0$ possess the same eigenvectors, and situations closely linked to the EPs are encountered due to the exact level
crossings of $\hat{H}_0+\lambda\hat{V}_0$. Our original Hamiltonian
(\ref{hat-H-lambda-take-1}) can be now redisplayed as
\be \label{hat-H-lambda-take-2}
   \hat{H}(\lambda) \; = \; \hat{H}_0 \; + \; \lambda \, \hat{V}_0 \; + \; \lambda \, \hat{V}_1 \mez .
\ee
Formula (\ref{hat-H-lambda-take-2}) motivates us to think of a slightly more general Hamiltonian
\be \label{hat-H-lambda-take-3}
   \hat{H}(\lambda,\delta) \; = \; \hat{H}_0 \; + \; \lambda \, \hat{V}_0 \; + \; \lambda \, \delta \, \hat{V}_1 \mez ;
\ee
where $\delta \in [0,1]$ serves as an auxiliary switching parameter of the $\lambda \, \hat{V}_1$ term. Importantly, one has
$\hat{H}(\lambda,0)=\hat{H}_0+\lambda\hat{V}_0$ and $\hat{H}(\lambda,1)=\hat{H}(\lambda)$ of Eq.~(\ref{hat-H-lambda-take-2}). Moreover,
the EPs of $\hat{H}(\lambda,\delta)$ of Eq.~(\ref{hat-H-lambda-take-3}) move continuously in the complex $\lambda$-plane when the real valued
switching control parameter $\delta$ is set to increase gradually from 0 to 1. If so, it seems natural to examine the possibility
of finding the equations of motion (EOM) governing the "flux" or "dynamical propagation" of the mentioned EPs along the
"time coordinate" $\delta \in [0,1]$. One may even anticipate that an explicit solution of such EOM (where the initial conditions at
$\delta=0$ are provided by the presumably known EPs of $\hat{H}(\lambda,0)=\hat{H}_0+\lambda\hat{V}_0$) would lead to finding
the desired EPs of $\hat{H}(\lambda)=\hat{H}(\lambda,1)$. It is the purpose of our present article to adequately explore both theoretical
and practical merits of the just sketched approach.

The just presented idea corresponds essentially to implementing adequately and systematically the nonhermitian perturbation theory
in the presence of EPs, where the perturbation is invoked by parameteric shift $\delta \mapsto \delta + {\rm d}\delta$. Let us mention
in this context that the merits of nonhermitian perturbation theory in the presence of an EP have been recently exploited e.g.~in
Ref.~\cite{EPs-examples-1} (see the corresponding Supplementary material). We also point out that our approach is intimately related
to the Dyson-Pechukas theory of level dynamics (see e.g.~Ref.~\cite{Wang}), which however has been pursued so far just within the framework
of hermitian Hamiltonian formalism.

The paper is organized as follows. Section 2 provides a self contained systematic fully explicit theoretical derivation of the sought EOM
for the EPs. As such, Section 2 represents the most important hardcore material to be communicated by the present article. The issue of choosing
appropriate initial conditions for the EOM is conveniently relegated to {\sl Appendix A}. Section 3 describes a conceptually simple yet certainly
nontrivial toy model, intended to serve as a relatively strict test of our obtained EOM. Symmetry of this toy model (resulting in simultaneous
existence of multiple EPs at particular values of $\lambda$) is discussed. Subsequently, we present in Section 3 an outcome of the numerical solution of our EOM
for the aforementioned toy model, in order to highlight suitability of our EOM method for practical computations of the EPs. Finally, Section 4
contains the concluding remarks.

\clearpage

\section{Mathematical formulation}

\subsection{Preliminaries}

Let us consider an $N$-by-$N$ complex symmetric\footnote{An extension of our considerations to general non-symmetric Hamiltonians is also
possible and relatively straightforward. However, in the present article we prefer to deal only with symmetric Hamiltonians for the sake
of maximum simplicity. } Hamiltonian matrix $\hat{H}(\lambda,\delta)$ depending upon two
parameters $\lambda \in {\mathbb C}$ and $\delta \in {\mathbb R}$. The Hamiltonian $\hat{H}(\lambda,\delta)$ is acting
in the linear space ${\mathbb C}^N$ of $N$-component ket (column) vectors
\be
   | v ) \; = \; \left( \matrix{ v_1 \cr v_2 \cr \bm\vdots \cr v_N } \right) \mez .
\ee
The associated bra (row) vectors are simply $( v | = ( v_1 \; v_2 \; \cdots \; v_N )$. The adequate scalar product
(the so called $c$-product) is defined by prescription
\be
   ( v | v'\n ) \; = \; \sum_{n=1}^N \, v_n \, v'_n \; = \; ( v'\n | v ) \mez .
\ee
Recall that the self overlap $(v|v)$ is generally complex valued, and $(v|v)=0$ does not imply $|v)$
equal to the zero vector $|\emptyset)$, see Chapter 9 of Ref.~\cite{NM-Book} for details.

Consistently with our motivational considerations outlined in the Introduction, we shall hereafter assume that there exists
a function $\lambda(\delta)$ such that, for each $\delta \in {\mathbb R}$, an eigenproblem of $\hat{H}(\lambda(\delta),\delta)$ gives rise to $M$ distinct binary EPs $[\,$$N \ge 2M$, each binary EP is formed via coalescence of two eigenvectors of
$\hat{H}(\lambda\to\lambda(\delta),\delta)$$\,]$.\footnote{
In the case when $\hat{H}(\lambda(\delta),\delta)$ does not possess any kind of symmetry, we expect $M=1$.
On the other hand, symmetries of $\hat{H}(\lambda(\delta),\delta)$ might imply $M>1$ (see Section 3 for an example).}
We define for later convenience
\be \label{H-delta-def}
   \hat{H}(\delta) \; \equiv \; \hat{H}(\lambda(\delta),\delta) \mez ;
\ee
and also
\be \label{V-delta-def}
   \hat{V}\m(\delta) \; \equiv \; {\rm d}_\delta \, \hat{H}(\delta) \; = \;
   \partial_\lambda \, \hat{H}(\lambda(\delta),\delta) \, \bm\dot{\lambda}(\delta) \; + \; \partial_\delta \, \hat{H}(\lambda(\delta),\delta) \mez ;
\ee
where $\bm\dot{\lambda}(\delta)={\rm d}_\delta\,\lambda(\delta)$, with ${\rm d}_\bullet=\frac{{\rm d}}{{\rm d}\bullet}$ and
$\partial_\bullet=\frac{\partial}{\partial\bullet}$.

The eigenproblem of our interest looks then as follows. In accordance with our above made assumption, for each
$\delta \in {\mathbb R}$ there exist $M$ binary EPs of $\hat{H}(\delta)$, satisfying
\be \label{H-delta-EVP-M-EPs}
   \hat{H}(\delta) \, | \tilde{c}_m^\delta ) \; = \; \tilde{E}_m^\delta \, | \tilde{c}_m^\delta ) \mez , \mez
   1 \leq m \leq M \mez ;
\ee
with obvious notations.\footnote{
The upper tilde superscript indicates here entities associated inherently with the EPs,
whereas all the non-EP entities are conveniently left without tilde. In this manner we distinguish e.g.~between
$\tilde{E}_1^\delta$ of Eq.~(\ref{H-delta-EVP-M-EPs}) and $E_1^\delta$ of Eq.~(\ref{H-delta-EVP-ordinary}). }
Besides these $M$ EPs, there exist also $(N-2M)$ ordinary non-degenerate non-EP eigenvectors of $\hat{H}(\delta)$, satisfying
\be \label{H-delta-EVP-ordinary}
   \hat{H}(\delta) \, | c_j^\delta ) \; = \; E_j^\delta \, | c_j^\delta ) \mez , \mez 1 \leq j \leq (N-2M) \mez .
\ee
Since the just listed ensemble of $(N-M)$ Hamiltonian eigenvectors $| \tilde{c}_m^\delta )$ and
$| c_j^\delta )$ does not form a complete basis set of ${\mathbb C}^N$\m, one needs to include into the game also $M$
complementary basis vectors (see Section 9.2 of Ref.~\cite{NM-Book}), satisfying
\be \label{H-delta-EVP-complementary}
   \left( \hat{H}(\delta) \, - \, \tilde{E}_m^\delta \, \hat{1} \right) \, | \tilde{b}_m^\delta ) \; = \; f_m^\delta \, | \tilde{c}_m^\delta ) \mez , \mez 1 \leq m \leq M \mez .
\ee
Here $f_m^\delta$ are nonzero coefficients arising from imposing suitable normalization conventions\footnote{
See equation (\ref{rescalings}) below and the accompanying discussion.
{\it Subsection 2.2} and {\sl Appendix A} describe an unambigous gauge fixing of $f_m^\delta$ and all related matters.}
% Rescale $| \tilde{c}_m^\delta )$ by any factor $\alpha_m^\delta \neq 0$, and $| \tilde{b}_m^\delta )$ by any factor $\beta_m^\delta \neq 0$,
% such that
% $| \tilde{c}_m^{\delta,{\rm new}} ) = (\alpha_m^\delta)^{-1}\,| \tilde{c}_m^\delta )$ and
% $| \tilde{b}_m^{\delta,{\rm new}} ) = (\beta_m^\delta)^{+1}\,| \tilde{b}_m^\delta )$. Then
% $\left( \hat{H}(\delta) \, - \, \tilde{E}_m^\delta \, \hat{1} \right) \, | \tilde{b}_m^{\delta,{\rm new}} )
% \, = \, f_m^{\delta,{\rm new}} \, | \tilde{c}_m^{\delta,{\rm new}} )$ with $f_m^{\delta,{\rm new}} \, = \,
% \alpha_m^\delta \, \beta_m^\delta \, f_m^\delta$. The overlap $( \tilde{b}_m^{\delta,{\rm new}} | \tilde{c}_m^{\delta,{\rm new}} ) =
% (\alpha_m^\delta)^{-1} \, (\beta_m^\delta)^{+1} \, ( \tilde{b}_m^{\delta} | \tilde{c}_m^{\delta} )$.
for $| \tilde{c}_m^\delta )$ and $| \tilde{b}_m^\delta )$, other notations are again self explanatory.

The corresponding orthonormality relations take the following explicit appearance:
\begin{eqnarray}
%  --------------------------------------------------------------------------------------------------------------------
   \label{onrel-c-j-c-j'} (c_j^\delta|c_{j'}^\delta) & = & \delta_{jj'} \mez ;\\
%  --------------------------------------------------------------------------------------------------------------------
   \label{onrel-c-j-c-m} (c_j^\delta|\tilde{c}_{m}^\delta) & = & 0 \mez ;\\
%  --------------------------------------------------------------------------------------------------------------------
   \label{onrel-c-j-b-m} (c_j^\delta|\tilde{b}_{m}^\delta) & = & 0 \mez ;\\
%  --------------------------------------------------------------------------------------------------------------------
   \label{onrel-c-m-c-m'} (\tilde{c}_{m}^\delta|\tilde{c}_{m'}^\delta) & = & 0 \mez ;\\
%  --------------------------------------------------------------------------------------------------------------------
   \label{onrel-c-m-b-m'} (\tilde{c}_{m}^\delta|\tilde{b}_{m'}^\delta) & = & \delta_{mm'} \mez ;\\
   %  --------------------------------------------------------------------------------------------------------------------
   \label{onrel-b-m-b-m'} (\tilde{b}_{m}^\delta|\tilde{b}_{m'}^\delta) & = & 0 \mez .
%  --------------------------------------------------------------------------------------------------------------------
\end{eqnarray}
Relations (\ref{onrel-c-m-c-m'}), (\ref{onrel-b-m-b-m'}) show that the eigenvectors $| \tilde{c}_m^\delta )$ and their complements $| \tilde{b}_m^\delta )$ are self orthogonal and normalized via (\ref{onrel-c-m-b-m'}) and (\ref{H-delta-EVP-complementary}), as opposed to the eigenvectors $| c_j^\delta )$ which are unit normalizable through (\ref{onrel-c-j-c-j'}). The pertinent closure property
is built up accordingly (see again Section 9.2 of Ref.~\cite{NM-Book}), we have
\be \label{closure}
   \sum_j \, | c_j^\delta ) ( c_j^\delta | \; + \;
   \sum_m \, | \tilde{c}_m^\delta ) ( \tilde{b}_m^\delta | \, + \,
   | \tilde{b}_m^\delta ) ( \tilde{c}_m^\delta | \; = \; \hat{1} \mez ;
\ee
where $\hat{1}$ stands for an $N$-by-$N$ unit matrix.

The normalization of self-orthogonal vectors $|\tilde{c}_{m}^\delta)$ and $|\tilde{b}_{m}^\delta)$ is not unambiguously fixed by
the formulas (\ref{H-delta-EVP-complementary}), (\ref{onrel-c-m-b-m'}), (\ref{closure}). Indeed, these relations
are invariant with respect to rescalings
\be \label{rescalings}
   \hspace*{-1.00cm}
   |\tilde{c}_{m}^{\delta,{\rm new}}) \; = \; g_m^\delta \; |\tilde{c}_{m}^\delta) \mz , \mz
   |\tilde{b}_{m}^{\delta,{\rm new}}) \; = \; \left(g_m^\delta\right)^{-1} |\tilde{b}_{m}^\delta) \mz , \mz
   f_m^{\delta,{\rm new}} \; = \; \left(g_m^\delta\right)^{-2} f_m^\delta \mz ;
\ee
where $g_m^\delta$ stands for any nozero complex valued factor.

Our above outlined formulas (\ref{H-delta-EVP-M-EPs})-(\ref{closure}) indicate that full solution of an eigenvalue
problem of $\hat{H}(\delta)$ is determined by seven fundamental entities
\be \label{entities}
   \lambda(\delta) \mz , \mz \tilde{E}_m^\delta \mz , \mz | \tilde{c}_m^\delta ) \mz , \mz | \tilde{b}_m^\delta ) \mz , \mz
   E_j^\delta \mz , \mz | c_j^\delta ) \mz , \mz f_m^\delta \mez .
\ee
The just displayed entities (\ref{entities}) depend continuously upon the parameter $\delta \in {\mathbb R}$.
An infinitesimal shift of $\delta$ changes our Hamiltonian (\ref{H-delta-def}) into
\be \label{H-perturbation}
   \hat{H}(\delta+{\rm d}\delta) \; = \; \hat{H}(\delta) \; + \; \hat{V}(\delta) \, {\rm d}\delta \mez .
\ee
This invokes the corresponding infinitesimal changes in the eigensolutions (\ref{entities}). The associated rates of change
\be \label{rates}
   \bm\dot{\lambda}(\delta) \mz , \mz \bm\dot{\tilde{E}_m^\delta} \mz , \mz | \bm\dot{\tilde{c}_m^\delta} ) \mz , \mz
   | \bm\dot{\tilde{b}_m^\delta} ) \mz , \mz \bm\dot{E}_j^\delta \mz , \mz | \bm\dot{c}_j^\delta ) \mz , \mz
   \bm\dot{f}_m^\delta
\ee
are obtainable by examining how do the eigensolutions (\ref{entities})
respond to the Hamiltonian perturbation $\hat{V}(\delta) \, {\rm d}\delta$ in equation (\ref{H-perturbation}). An explicit
analytic elaboration of such a perturbation theory is by no means conventional or trivial, since the considered eigenproblems
of $\hat{H}(\delta)$ and $\hat{H}(\delta+{\rm d}\delta)$ do support $M$ binary EPs, as highlighted above in
(\ref{H-delta-EVP-M-EPs})-(\ref{closure}). Nevertheless, the just mentioned task is feasible to perform,
and results in explicit analytic prescriptions for the "velocities" (\ref{rates}) determining the "dynamics" or "motion"
of the seven fundamental eigensolution entities (\ref{entities}) in the flux of "time" $\delta \in {\mathbb R}$.
These {\sl "equations of motion for the EPs"} (or briefly EOM) are worked out in a self contained manner in the next {\it Subsection 2.2}, which
actually represents the most important "hard core" material to be communicated by the present paper. See the resulting equations
(\ref{dot-lambda-EOM}), (\ref{dot-E-m-final}), (\ref{dot-E-j-final}), (\ref{dot-f-m-EOM}), (\ref{dot-c-m-EOM}),
(\ref{dot-c-j-EOM}), (\ref{dot-b-m-EOM}) below. Furthermore, an additional {\sl Appendix A} describes in a self contained fashion
the construction of adequate initial conditions for these EOM, corresponding to a frequently encountered situation when the sought EPs
emanate from hermitian curve crossings/degeneracies of $\hat{H}(\lambda,\delta)$.

\subsection{Equations of motion for the exceptional points}

Assume that the seven fundamental entities (\ref{entities}) are known for a given $\delta \in {\mathbb R}$.
Let us derive now in a self contained manner explicit analytic formulas for the corresponding (presumably unknown)
derivatives (\ref{rates}). These need to be expressed solely in terms of the known quantities (\ref{entities}).

\subsubsection{An equation of motion for $\bm\dot{\lambda}(\delta)$}

\phantom{bla}
\vspace*{+0.20cm}

Take equation (\ref{H-delta-EVP-M-EPs}) for a given value of $m$ ($1 \leq m \leq M$). Differentiate both sides with
respect to $\delta$, as to get
\be \label{H-delta-EVP-M-EPs-dot}
   \hat{V}(\delta) \, | \tilde{c}_m^\delta ) \; + \; \hat{H}(\delta) \, | \bm\dot{\tilde{c}_m^\delta} ) \; = \;
   \bm\dot{\tilde{E}_m^\delta} \, | \tilde{c}_m^\delta ) \; + \; \tilde{E}_m^\delta \, | \bm\dot{\tilde{c}_m^\delta} ) \mez .
\ee
Substitute (\ref{V-delta-def}), multiply subsequently by $( \tilde{c}_m^\delta |$ from the left, and
exploit the self orthogonality property (\ref{onrel-c-m-c-m'}) at $m'\n=m$ which implies also
\be \label{overlap-1}
   ( \tilde{c}_m^\delta | \bm\dot{\tilde{c}_m^\delta} ) \; = \; 0 \mez .
\ee
This yields a compelling formula
\be \label{dot-lambda-EOM}
   \bm\dot{\lambda}(\delta) \; = \; - \, \frac{(\tilde{c}_m^\delta| \partial_\delta \, \hat{H}(\lambda(\delta),\delta) |\tilde{c}_m^\delta)}
   {(\tilde{c}_m^\delta| \partial_\lambda \, \hat{H}(\lambda(\delta),\delta) |\tilde{c}_m^\delta)} \mez ;
\ee
which represents perhaps the most important result of the present paper. Outcome (\ref{dot-lambda-EOM}) should be regarged
as an equation of motion for $\bm\dot{\lambda}(\delta)$. The r.h.s.~of (\ref{dot-lambda-EOM}) must be
independent upon $m$, as long as our assumption of having $M$ binary EPs holds. The $m$-independence of (\ref{dot-lambda-EOM}) serves as an useful check of internal consistency in our numerical calculations of Section 3 of the main text.
From now on, $\bm\dot{\lambda}(\delta)$ will be regarded as explicitly known (and presumably finite\footnote{It is beyond the scope of the present
article to examine if (or under which circumstances) the denominator of (\ref{dot-lambda-EOM}) can ever become zero.}),
and the same applies also for the perturbation $\hat{V}\m(\delta)$ of Eq.~(\ref{V-delta-def}).

\subsubsection{Equations of motion for $\bm\dot{\tilde{E}_m^\delta}$, $\bm\dot{E}_j^\delta$, and $\bm\dot{f}_m^\delta$,
plus other accompanying elaborations}

\phantom{bla}
\vspace*{+0.20cm}

Take equation (\ref{H-delta-EVP-M-EPs-dot}) and multiply from the left by $( \tilde{c}_{m'}^\delta\n |$ where $m' \n\neq m$. Exploit subsequently (\ref{onrel-c-m-c-m'}). This yields an overlap element
\be \label{overlap-2}
   ( \tilde{c}_{m'}^\delta\n | \bm\dot{\tilde{c}_m^\delta} ) \; = \;
   \frac{( \tilde{c}_{m'}^\delta | \hat{V}\m(\delta) | \tilde{c}_m^\delta )}{\tilde{E}_m^\delta-\tilde{E}_{m'}^\delta}
   \mez . \mez [\,m'\n \neq m\,]
\ee
The denominator of (\ref{overlap-2}) is nonsingular as long as the considered $M$ binary EPs are distinct.

Take again (\ref{H-delta-EVP-M-EPs-dot}) and multiply from the left by $( \tilde{b}_m^\delta\n |$. Exploit subsequently
(\ref{H-delta-EVP-complementary}), (\ref{overlap-1}), and also (\ref{onrel-c-m-b-m'}) for $m'\n=m$. This yields the as yet unknown energy derivative
\be \label{dot-E-m-final}
   \bm\dot{\tilde{E}_m^\delta} \; = \; ( \tilde{b}_m^\delta | \hat{V}\m(\delta) | \tilde{c}_m^\delta ) \mez .
\ee
This is the sought equation of motion for $\bm\dot{\tilde{E}_m^\delta}$.

Take again (\ref{H-delta-EVP-M-EPs-dot}) and multiply from the left by $( \tilde{b}_{m'}^\delta\n |$ where $m' \n\neq m$. Exploit subsequently (\ref{H-delta-EVP-complementary}) and (\ref{onrel-c-m-b-m'}) together with (\ref{overlap-2}).
This yields an overlap element
\be \label{overlap-3}
   ( \tilde{b}_{m'}^\delta\n | \bm\dot{\tilde{c}_m^\delta} ) \; = \;
   \frac{( \tilde{b}_{m'}^\delta | \hat{V}\m(\delta) | \tilde{c}_m^\delta )}{\tilde{E}_m^\delta-\tilde{E}_{m'}^\delta}
   \; + \; f_{m'}^\delta \, \frac{( \tilde{c}_{m'}^\delta | \hat{V}\m(\delta) | \tilde{c}_m^\delta )}
   {(\tilde{E}_m^\delta-\tilde{E}_{m'}^\delta)^2} \mez . \mez [\,m'\n \neq m\,]
\ee
The denominators are nonsingular for the same reason as in (\ref{overlap-2}). In passing we note that
\be \label{overlap-3b}
   ( \tilde{c}_m^\delta | \bm\dot{\tilde{b}_{m'}^\delta\n} ) \; = \;
   - \, ( \tilde{b}_{m'}^\delta\n | \bm\dot{\tilde{c}_m^\delta} ) \mez ;
\ee
valid as an immediate consequence of (\ref{onrel-c-m-b-m'}).

Take again (\ref{H-delta-EVP-M-EPs-dot}) and multiply from the left by $( c_j^\delta |$ where $1 \leq j \leq (N-2\,M)$.
Exploit subsequently (\ref{H-delta-EVP-ordinary}) and (\ref{onrel-c-j-c-m}). This yields an overlap element
\be \label{overlap-4}
   ( c_j^\delta\n | \bm\dot{\tilde{c}_m^\delta} ) \; = \;
   \frac{( c_j^\delta | \hat{V}\m(\delta) | \tilde{c}_m^\delta )}{\tilde{E}_m^\delta-E_j^\delta} \mez .
\ee
The denominator is nonsingular as long as the considered $m$-th EP eigenvalue $\tilde{E}_m^\delta$ does not coincide with the non-EP eigenvalues $E_j$.

Proceeding further, take equation (\ref{H-delta-EVP-ordinary}) for a given value of $j$ ($1 \leq j \leq N-2\,M$). Differentiate both sides with respect to $\delta$, as to get
\be \label{H-delta-EVP-ordinary-dot}
   \hat{V}(\delta) \, | c_j^\delta ) \; + \; \hat{H}(\delta) \, | \bm\dot{c}_j^\delta ) \; = \;
   \bm\dot{E}_j^\delta \, | c_j^\delta ) \; + \; E_j^\delta \, | \bm\dot{c}_j^\delta ) \mez .
\ee
Multiply subsequently from the left by $( c_j^\delta |$, exploit then (\ref{H-delta-EVP-ordinary}) and
(\ref{onrel-c-j-c-j'}) for $j'\n=j$. This yields the as yet unknown energy derivative
\be \label{dot-E-j-final}
   \bm\dot{E}_j^\delta \; = \; ( c_j^\delta | \hat{V}\m(\delta) | c_j^\delta ) \mez .
\ee
This is the sought equation of motion for $\bm\dot{E}_j^\delta$.

Take again (\ref{H-delta-EVP-ordinary-dot}) and multiply from the left by $( c_{j'}^\delta |$ where $j'\n \neq j$.
Exploit then (\ref{H-delta-EVP-ordinary}) and (\ref{onrel-c-j-c-j'}). This yields an overlap element
\be \label{overlap-5}
   ( c_{j'}^\delta | \bm\dot{c}_j^\delta ) \; = \;
   \frac{( c_{j'}^\delta | \hat{V}\m(\delta) | c_j^\delta )}{E_j^\delta-E_{j'}^\delta} \mez . \mez [\,j'\n \neq j\,]
\ee
The denominator is nonsingular as long as the non-EP eigenvalues are non-degenerate.

Take again (\ref{H-delta-EVP-ordinary-dot}) and multiply from the left by $( \tilde{c}_m^\delta |$.
Exploit then (\ref{H-delta-EVP-M-EPs}) and (\ref{onrel-c-j-c-m}). This yields an overlap element
\be \label{overlap-6}
   ( \tilde{c}_m^\delta | \bm\dot{c}_j^\delta ) \; = \;
   \frac{( \tilde{c}_m^\delta | \hat{V}\m(\delta) | c_j^\delta )}{E_j^\delta-\tilde{E}_m^\delta} \mez .
\ee
The denominator is nonsingular as long as $E_j$ does not coincide with the EP eigenvalues $\tilde{E}_m^\delta$.

Take again (\ref{H-delta-EVP-ordinary-dot}) and multiply from the left by $( \tilde{b}_m^\delta |$.
Exploit then (\ref{H-delta-EVP-complementary}) and (\ref{onrel-c-j-b-m}) together with (\ref{overlap-6}).
This yields an overlap element
\be \label{overlap-7}
   ( \tilde{b}_m^\delta | \bm\dot{c}_j^\delta ) \; = \; \frac{( \tilde{b}_m^\delta | \hat{V}\m(\delta) | c_j^\delta )}{E_j^\delta-\tilde{E}_m^\delta} \; + \; f_m^\delta \, \frac{( \tilde{c}_m^\delta | \hat{V}\m(\delta) | c_j^\delta )}{(E_j^\delta-\tilde{E}_m^\delta)^2} \mez .
\ee
Again, the denominators are nonsingular as long as $E_j$ does not coincide with the EP eigenvalues $\tilde{E}_m^\delta$.
In passing we note that
\be \label{overlap-7b}
   ( c_j^\delta | \bm\dot{\tilde{b}_{m}^\delta} ) \; = \;
   - \, ( \tilde{b}_m^\delta | \bm\dot{c}_j^\delta ) \mez ;
\ee
valid as an immediate consequence of (\ref{onrel-c-j-b-m}).

Proceeding further, take equation (\ref{H-delta-EVP-complementary}) for a given value of $m$ ($1 \leq m \leq M$). Differentiate both sides with respect to $\delta$, as to get
\be \label{H-delta-EVP-complementary-dot}
   \hspace*{-1.50cm}
   \hat{V}(\delta) \, | \tilde{b}_m^\delta ) \; + \; \hat{H}(\delta) \, | \bm\dot{\tilde{b}_m^\delta} ) \; - \;
   \bm\dot{\tilde{E}_m^\delta} \, | \tilde{b}_m^\delta ) \; - \; \tilde{E}_m^\delta \, | \bm\dot{\tilde{b}_m^\delta} )
   \; = \; \bm\dot{f}_m^\delta \, | \tilde{c}_m^\delta ) \; + \; f_m^\delta \, | \bm\dot{\tilde{c}_m^\delta} ) \mez .
\ee
Multiply subsequently from the left by $( \tilde{b}_m^\delta |$, exploit then (\ref{H-delta-EVP-complementary}) and
(\ref{onrel-c-m-b-m'}) for $m'\n=m$, as well as (\ref{onrel-b-m-b-m'}) for $m'\n=m$. This yields another important relation
\be \label{dot-f-m-intermediate}
   ( \tilde{b}_m^\delta | \hat{V}(\delta) \, | \tilde{b}_m^\delta ) \; = \; \bm\dot{f}_m^\delta \; + \;
   f_m^\delta \, ( \tilde{b}_m^\delta | \bm\dot{\tilde{c}_m^\delta} ) \; - \;
   f_m^\delta \, ( \tilde{c}_m^\delta | \bm\dot{\tilde{b}_m^\delta} ) \mez .
\ee

Take again (\ref{H-delta-EVP-complementary-dot}) and multiply from the left by $( \tilde{b}_{m'}^\delta\n |$
where $m' \n\neq m$. Exploit subsequently (\ref{H-delta-EVP-complementary}) and (\ref{onrel-c-m-b-m'}),
(\ref{onrel-b-m-b-m'}), together with (\ref{overlap-3}) and (\ref{overlap-3b}). This yields an overlap element
\begin{eqnarray} \label{overlap-8}
%  ----------------------------------------------------------------------------------------------------------------------
   \hspace*{-3.00cm} & & ( \tilde{b}_{m'}^\delta | \bm\dot{\tilde{b}_m^\delta} ) \; =  \hspace*{+10.00cm} [\,m'\n \neq m\,]\\
%  ----------------------------------------------------------------------------------------------------------------------
   \hspace*{-3.00cm} & = & \frac{( \tilde{b}_{m'}^\delta | \hat{V}\m(\delta) | \tilde{b}_m^\delta )}
   {\tilde{E}_m^\delta-\tilde{E}_{m'}^\delta} \; + \; f_{m'}^\delta \;
   \frac{( \tilde{c}_{m'}^\delta | \hat{V}\m(\delta) | \tilde{b}_m^\delta )}
   {(\tilde{E}_m^\delta-\tilde{E}_{m'}^\delta)^2}
   \; - \; f_m^\delta \;
   \frac{( \tilde{b}_{m'}^\delta | \hat{V}\m(\delta) | \tilde{c}_m^\delta )}
   {(\tilde{E}_m^\delta-\tilde{E}_{m'}^\delta)^2}
   \; - \; 2 \, f_m^\delta \, f_{m'}^\delta \;
   \frac{( \tilde{c}_{m'}^\delta | \hat{V}\m(\delta) | \tilde{c}_m^\delta )}
   {(\tilde{E}_m^\delta-\tilde{E}_{m'}^\delta)^3} \mz . \nonumber
%  ----------------------------------------------------------------------------------------------------------------------
\end{eqnarray}
Much like pointed out before, the denominator is nonsingular as long as the considered $M$ binary EPs are distinct.

Take again (\ref{H-delta-EVP-complementary-dot}) and multiply from the left by $( \tilde{c}_{m}^\delta\n |$.
Exploit subsequently (\ref{H-delta-EVP-M-EPs}), (\ref{onrel-c-m-c-m'}) for $m=m'$\m, (\ref{onrel-c-m-b-m'}) for $m=m'$\m,
and (\ref{overlap-1}). This yields $\bm\dot{\tilde{E}_m^\delta}=( \tilde{c}_m^\delta | \hat{V}\m(\delta) | \tilde{b}_m^\delta )$ as already known from (\ref{dot-E-m-final}).

Take again (\ref{H-delta-EVP-complementary-dot}) and multiply from the left by $( \tilde{c}_{m'}^\delta\n |$
where $m' \n\neq m$. Exploit subsequently (\ref{H-delta-EVP-M-EPs}), (\ref{onrel-c-m-c-m'}), (\ref{onrel-c-m-b-m'}),
together with (\ref{overlap-2}). This yields an overlap element $( \tilde{c}_{m'}^\delta\n | \bm\dot{\tilde{b}_m^\delta} )$
in the form exactly equal to an immediate consequence of (\ref{overlap-3}) and (\ref{overlap-3b}).

Take again (\ref{H-delta-EVP-complementary-dot}) and multiply from the left by $( c_j^\delta\n |$ where $1 \leq j \leq (N-2\,M)$. Exploit subsequently (\ref{H-delta-EVP-ordinary}), (\ref{onrel-c-j-c-m}), (\ref{onrel-c-j-b-m}), together with
(\ref{overlap-4}). This yields an overlap element $( c_j^\delta | \bm\dot{\tilde{b}_m^\delta} )$ in the form exactly equal
to an immediate consequence of (\ref{overlap-7}) and (\ref{overlap-7b}).

To complete all our technical elaborations regarding the overlap elements, we need to specify the as yet
undetermined quantities $( c_j^\delta | \bm\dot{c}_j^\delta )$, $( \tilde{b}_m^\delta | \bm\dot{\tilde{b}_m^\delta} )$,
$( \tilde{b}_m^\delta | \bm\dot{\tilde{c}_m^\delta} )$, $( \tilde{c}_m^\delta | \bm\dot{\tilde{b}_m^\delta} )$.
Recall that $( \tilde{c}_m^\delta | \bm\dot{\tilde{c}_m^\delta} )$ is already fixed by (\ref{overlap-1}). Clearly,
property (\ref{onrel-c-j-c-j'}) for $j'\n=j$ implies immediately
\be \label{overlap-9}
   ( c_j^\delta | \bm\dot{c}_j^\delta ) \; = \; 0 \mez .
\ee
Similarly, property (\ref{onrel-b-m-b-m'}) for $m'\n=m$ yields immediately
\be \label{overlap-10}
   ( \tilde{b}_m^\delta | \bm\dot{\tilde{b}_m^\delta} ) \; = \; 0 \mez .
\ee
An appropriate discussion of $( \tilde{b}_m^\delta | \bm\dot{\tilde{c}_m^\delta} )$ and
$( \tilde{c}_m^\delta | \bm\dot{\tilde{b}_m^\delta} )$ is a bit more intriguing. Property
(\ref{onrel-c-m-b-m'}) for $m'\n=m$ yields immediately
\be \label{overlap-11}
   ( \tilde{b}_m^\delta | \bm\dot{\tilde{c}_m^\delta} ) \; = \; - \, ( \tilde{c}_m^\delta | \bm\dot{\tilde{b}_m^\delta} )
   \mez ;
\ee
hence it is sufficient to determine just $( \tilde{b}_m^\delta | \bm\dot{\tilde{c}_m^\delta} )$. Importantly, the
self-orthogonal vectors $| \tilde{c}_m^\delta )$ and $| \tilde{b}_m^\delta )$, as well as the factor $f_m^\delta$,
have been introduced in the main text only modulo the rescaling transformation (\ref{rescalings}). It is a trivial
matter to verify that the as yet arbitrary rescaling coefficients $g_m^\delta$ can be always chosen in such a particular
manner as to arrange for having
\be \label{overlap-12}
   ( \tilde{b}_m^\delta | \bm\dot{\tilde{c}_m^\delta} ) \; = \; 0 \; = \;
   ( \tilde{c}_m^\delta | \bm\dot{\tilde{b}_m^\delta} ) \mez .
\ee
This is our suitably chosen gauge fixing convention for $( \tilde{b}_m^\delta | \bm\dot{\tilde{c}_m^\delta} )$ and
$( \tilde{c}_m^\delta | \bm\dot{\tilde{b}_m^\delta} )$. Having imposed (\ref{overlap-12}), equation
(\ref{dot-f-m-intermediate}) simplifies into a finalized equation of motion for $\bm\dot{f}_m^\delta$, namely,
\be \label{dot-f-m-EOM}
   \bm\dot{f}_m^\delta \; = \; ( \tilde{b}_m^\delta | \hat{V}\m(\delta) | \tilde{b}_m^\delta ) \mez .
\ee
For the sake of completeness and clarity, let us also point out here that
$( \tilde{c}_m^\delta | \hat{V}\m(\delta) | \tilde{c}_m^\delta ) = 0$, this is equivalent to (\ref{dot-lambda-EOM}).

\subsubsection{Equations of motion for $| \bm\dot{\tilde{c}_m^\delta} )$, $| \bm\dot{c}_j^\delta )$, $| \bm\dot{\tilde{b}_m^\delta} )$}

\phantom{bla}
\vspace*{+0.20cm}

The closure property (\ref{closure}) combined with (\ref{overlap-1}), (\ref{overlap-2}), (\ref{overlap-3}), (\ref{overlap-4}), (\ref{overlap-12}) provides immediately the desired equation of motion for
$| \bm\dot{\tilde{c}_m^\delta} )$. One has
\begin{eqnarray} \label{dot-c-m-EOM}
%  --------------------------------------------------------------------------------------------------------------------
   | \bm\dot{\tilde{c}_m^\delta} ) & = & \sum_{j} \, | c_j^\delta ) \;
   \frac{( c_j^\delta | \hat{V}\m(\delta) | \tilde{c}_m^\delta )}{\tilde{E}_m^\delta-E_j^\delta} \nonumber\\
%  --------------------------------------------------------------------------------------------------------------------
   & + & \sum_{m' \neq m} \, | \tilde{c}_{m'}^\delta ) \, \frac{( \tilde{b}_{m'}^\delta | \hat{V}\m(\delta) | \tilde{c}_m^\delta )}{\tilde{E}_m^\delta-\tilde{E}_{m'}^\delta} \; + \; \sum_{m' \neq m} \, | \tilde{c}_{m'}^\delta ) \,
   f_{m'}^\delta \, \frac{( \tilde{c}_{m'}^\delta | \hat{V}\m(\delta) | \tilde{c}_m^\delta )}{(\tilde{E}_m^\delta-\tilde{E}_{m'}^\delta)^2} \nonumber\\
%  --------------------------------------------------------------------------------------------------------------------
   & + & \sum_{m' \neq m} \, | \tilde{b}_{m'}^\delta ) \, \frac{( \tilde{c}_{m'}^\delta | \hat{V}\m(\delta) | \tilde{c}_m^\delta )}{\tilde{E}_m^\delta-\tilde{E}_{m'}^\delta} \mez .
%  --------------------------------------------------------------------------------------------------------------------
\end{eqnarray}

Similarly, the closure property (\ref{closure}) combined with (\ref{overlap-5}), (\ref{overlap-6}), (\ref{overlap-7}), (\ref{overlap-9}) provides immediately the desired equation of motion for $| \bm\dot{c}_j^\delta )$. One has
\begin{eqnarray} \label{dot-c-j-EOM}
%  --------------------------------------------------------------------------------------------------------------------
   | \bm\dot{c}_j^\delta ) & = & \sum_{j' \neq j} \, | c_{j'}^\delta ) \, \frac{( c_{j'}^\delta | \hat{V}\m(\delta) | c_j^\delta )}{E_j^\delta-E_{j'}^\delta} \nonumber\\
%  --------------------------------------------------------------------------------------------------------------------
   & + & \sum_m \, | \tilde{c}_m^\delta ) \, \frac{( \tilde{b}_m^\delta | \hat{V}\m(\delta) | c_j^\delta )}{E_j^\delta-\tilde{E}_m^\delta} \; + \;
   \sum_m \, | \tilde{c}_m^\delta ) \, f_m^\delta \, \frac{( \tilde{c}_m^\delta | \hat{V}\m(\delta) | c_j^\delta )}{(E_j^\delta-\tilde{E}_m^\delta)^2} \nonumber\\
%  --------------------------------------------------------------------------------------------------------------------
   & + & \sum_m \, | \tilde{b}_m^\delta ) \, \frac{( \tilde{c}_m^\delta | \hat{V}\m(\delta) | c_j^\delta )}{E_j^\delta-\tilde{E}_m^\delta} \mez .
%  --------------------------------------------------------------------------------------------------------------------
\end{eqnarray}

Finally, the closure property (\ref{closure}) combined with (\ref{overlap-3}), (\ref{overlap-3b}), (\ref{overlap-7}), (\ref{overlap-7b}), (\ref{overlap-8}), (\ref{overlap-10}), (\ref{overlap-12}) provides immediately the desired equation
of motion for $| \bm\dot{\tilde{b}_m^\delta} )$. One has
\begin{eqnarray} \label{dot-b-m-EOM}
%  --------------------------------------------------------------------------------------------------------------------
   \hspace*{-2.00cm} & & | \bm\dot{\tilde{b}_m^\delta} ) \; = \nonumber\\
   \hspace*{-2.00cm} & = & \sum_{j} \, | c_j^\delta ) \; \frac{( c_j^\delta | \hat{V}\m(\delta) | \tilde{b}_m^\delta )}
   {\tilde{E}_m^\delta-E_j^\delta} \; - \;
   f_m^\delta \, \sum_{j} \, | c_j^\delta ) \; \frac{( c_j^\delta | \hat{V}\m(\delta) | \tilde{c}_m^\delta )}
   {(\tilde{E}_m^\delta-E_j^\delta)^2} \nonumber\\
%  --------------------------------------------------------------------------------------------------------------------
   \hspace*{-2.00cm} & + & \sum_{m' \neq m} \, | \tilde{c}_{m'}^\delta ) \;
   \frac{( \tilde{b}_{m'}^\delta | \hat{V}\m(\delta) | \tilde{b}_m^\delta )}
   {\tilde{E}_m^\delta-\tilde{E}_{m'}^\delta} \nonumber\\
%  --------------------------------------------------------------------------------------------------------------------
   \hspace*{-2.00cm} & + & \sum_{m' \neq m} \, | \tilde{c}_{m'}^\delta ) \; f_{m'}^\delta \;
   \frac{( \tilde{c}_{m'}^\delta | \hat{V}\m(\delta) | \tilde{b}_m^\delta )}
   {(\tilde{E}_m^\delta-\tilde{E}_{m'}^\delta)^2}
   \; - \; \sum_{m' \neq m} \, | \tilde{c}_{m'}^\delta ) \; f_m^\delta \;
   \frac{( \tilde{b}_{m'}^\delta | \hat{V}\m(\delta) | \tilde{c}_m^\delta )}
   {(\tilde{E}_m^\delta-\tilde{E}_{m'}^\delta)^2} \nonumber\\
   \hspace*{-2.00cm} & & \hspace*{+5.20cm}
   \; - \; \sum_{m' \neq m} \, | \tilde{c}_{m'}^\delta ) \; 2 \, f_m^\delta \, f_{m'}^\delta \;
   \frac{( \tilde{c}_{m'}^\delta | \hat{V}\m(\delta) | \tilde{c}_m^\delta )}
   {(\tilde{E}_m^\delta-\tilde{E}_{m'}^\delta)^3} \nonumber\\
%  --------------------------------------------------------------------------------------------------------------------
   \hspace*{-2.00cm} & + & \sum_{m' \neq m} \, | \tilde{b}_{m'}^\delta ) \;
   \frac{( \tilde{c}_{m'}^\delta | \hat{V}\m(\delta) | \tilde{b}_m^\delta )}{\tilde{E}_m^\delta-\tilde{E}_{m'}^\delta}
   \; - \; f_m^\delta \, \sum_{m' \neq m} \, | \tilde{b}_{m'}^\delta ) \;
   \frac{( \tilde{c}_{m'}^\delta | \hat{V}\m(\delta) | \tilde{c}_m^\delta )}
   {(\tilde{E}_m^\delta-\tilde{E}_{m'}^\delta)^2} \mez .
%  --------------------------------------------------------------------------------------------------------------------
\end{eqnarray}

In summary, we have in hand now a self contained collection of seven mutually coupled equations of motion (\ref{dot-lambda-EOM}),
(\ref{dot-E-m-final}), (\ref{dot-E-j-final}), (\ref{dot-f-m-EOM}), (\ref{dot-c-m-EOM}), (\ref{dot-c-j-EOM}),
(\ref{dot-b-m-EOM}) for the derivatives (\ref{rates}). These EOM determine the flux of the seven fundamental
entities (\ref{entities}) in the "time" $\delta$, and can be propagated numerically
along $\delta \in {\mathbb R}$ once appropriate inital conditions are specified.
As already pointed out above, the issue of initial conditions is addressed in {\sl Appendix A}.

\section{Test in a simple toy model}

\subsection{Introducing the toy model}

The general mathematical formalism introduced above in Section 2 (and supplemented by {\sl Appendix A})
will be tested below on a conceptually simple yet quite nontrivial toy model. We deliberately choose here a quantum system
whose state space is finite dimensional, and which therefore gives rise to a finite number of EPs.
Moreover, an inherent symmetry of our toy model allows simultaneous existence of mutiple EPs ($M>1$), as anticipated already in Section 2, and as explained
in detail in the Figures below.

Our considered toy model corresponds to a theory of two distinct mutually coupled angular momenta,
$\hat{\vec{I}}=(\hat{I}_1,\hat{I}_2,\hat{I}_3)$ and $\hat{\vec{J}}=(\hat{J}_1,\hat{J}_2,\hat{J}_3)$,
which possess the conventional commutation properties. The associated starting
Hamiltonian is defined through prescription
\be \label{hat-H-lambda-1-def}
   \hat{H}(\lambda) \; = \; \hat{H}_0 \; + \; \lambda \, \hat{V} \mez ;
\ee
with
\be
   \hat{H}_0 \; = \; \omega \, \Bigl( \hat{I}_3 + \hat{J}_3 \Bigr) \mez ;
\ee
and
\be
   \hat{V} \; = \; \hat{I}_+ \, \hat{J}_- \, + \, \hat{I}_- \, \hat{J}_+ \, + \,
   \hat{I}_+ \, \hat{J}_+ \, + \, \hat{I}_- \, \hat{J}_- \; = \; 4 \, \hat{I}_1 \, \hat{J}_1 \mez .
\ee
Here $\omega>0$ and $\lambda \in {\mathbb C}$, and of course
\be
   \hat{I}_\pm \; = \; \hat{I}_1 \pm i\,\hat{I}_2 \mez , \mez
   \hat{J}_\pm \; = \; \hat{J}_1 \pm i\,\hat{J}_2 \mez .
\ee
The pertinent state space is spanned by basis vectors
$| \,I_{\rm T}\,I_3\,J_{\rm T}\,J_3\, \ra$, where $(I_{\rm T}(I_{\rm T}+1),I_3)$ are eigenvalues of
$(\hat{I}^2,\hat{I}_3)$ and similarly $(J_{\rm T}(J_{\rm T}+1),J_3)$ are eigenvalues of $(\hat{J}^2,\hat{J}_3)$.
Clearly, both $I_{\rm T}$ and $J_{\rm T}$
are good quantum numbers for the Hamiltonian (\ref{hat-H-lambda-1-def}). Dimension of a particular $(I_{\rm T},J_{\rm T})$ sector equals to
${\cal N}_I\,{\cal N}_J$, here ${\cal N}_I=2\,I_{\rm T}+1$, and similarly for ${\cal N}_J$. Hereafter we shall assume for definiteness
$I_{\rm T}=\frac{N}{2}$ and $J_{\rm T}=\frac{1}{2}$, where $N$ is an odd positive integer (correspondingly, ${\cal N}_I=N+1$ and ${\cal N}_J=2$).
The parity of $(I_3+J_3)$ is then another good quantum number.

Our primary interest consists in finding all the EPs of $\hat{H}(\lambda)$ of Eq.~(\ref{hat-H-lambda-1-def}) in the complex
$\lambda$-plane. To accomplish this goal, we follow the general strategy outlined in the Introduction, include into the game
an auxiliary switching parameter $\delta \in [0,1]$, and focus on investigating the parameterically $\delta$-dependent EPs of an
augmented Hamiltonian (\ref{hat-H-lambda-take-3}), where by definition
\be
   \hat{V}_0 \; = \; \hat{I}_+ \, \hat{J}_- \, + \, \hat{I}_- \, \hat{J}_+ \mez ;
\ee
\be
   \hat{V}_1 \; = \; \hat{I}_+ \, \hat{J}_+ \, + \, \hat{I}_- \, \hat{J}_- \mez .
\ee
Written explicitly, we have
\be \label{H-lambda-delta-def}
   \hspace*{-1.00cm}
   \hat{H}(\lambda,\delta) \; = \; \omega \, \Bigl( \hat{I}_3 + \hat{J}_3 \Bigr) \; + \; \lambda \, \left\{
   \hat{I}_+ \, \hat{J}_- \, + \, \hat{I}_- \, \hat{J}_+ \, + \, \delta \,
   \Bigl( \hat{I}_+ \, \hat{J}_+ \, + \, \hat{I}_- \, \hat{J}_- \Bigr) \right\} \mz .
\ee
Again, both $I_{\rm T}$ and $J_{\rm T}$ are good quantum numbers for the Hamiltonian (\ref{H-lambda-delta-def}),
and the parity of $(I_3+J_3)$ is another good quantum number.

Before proceeding further, let us highlight an additional symmetry of the Hamiltonian $\hat{H}(\lambda,\delta)$ of Eq.~(\ref{H-lambda-delta-def}).
The eigenvalue spectrum of $\hat{H}(\lambda,\delta)$ is clearly invariant with respect to any similarity
transformation. One such particular transformation is represented by an unitary operator
\be
   \hat{U} \; = \; e^{-i\pi\hat{I}_1} \; e^{-i\pi\hat{J}_2} \mez ;
\ee
which corresponds to rotation of $\hat{\vec{I}}$ by angle $\pi$ around the first coordinate axis, and to rotation of $\hat{\vec{J}}$
by angle $\pi$ around the second coordinate axis. Direct calculation yields
\be
   \hat{U}^\dagger \, \hat{I}_1 \, \hat{U} \; = \; + \, \hat{I}_1 \mez , \mez
   \hat{U}^\dagger \, \hat{I}_2 \, \hat{U} \; = \; - \, \hat{I}_2 \mez , \mez
   \hat{U}^\dagger \, \hat{I}_3 \, \hat{U} \; = \; - \, \hat{I}_3 \mez ;
\ee
and similarly
\be
   \hat{U}^\dagger \, \hat{J}_1 \, \hat{U} \; = \; - \, \hat{J}_1 \mez , \mez
   \hat{U}^\dagger \, \hat{J}_2 \, \hat{U} \; = \; + \, \hat{J}_2 \mez , \mez
   \hat{U}^\dagger \, \hat{J}_3 \, \hat{U} \; = \; - \, \hat{J}_3 \mez .
\ee
Hence
\be
   \hat{U}^\dagger \, \hat{I}_\pm \, \hat{U} \; = \; +\,\hat{I}_\mp \mez , \mez
   \hat{U}^\dagger \, \hat{J}_\pm \, \hat{U} \; = \; -\,\hat{J}_\mp \mez .
\ee
If so, then our Hamiltonian $\hat{H}(\lambda,\delta)$ of Eq.~(\ref{H-lambda-delta-def}) is converted into
\be \label{U-dagger-H-U}
   \hat{U}^\dagger \, \hat{H}(\lambda,\delta) \, \hat{U} \; = \; -\,\hat{H}(\lambda,\delta) \mez .
\ee
The just derived symmetry property (\ref{U-dagger-H-U}) reveals that both $\hat{H}(\lambda,\delta)$ and $-\hat{H}(\lambda,\delta)$ must possess
the same spectrum. Thus, if $E(\lambda,\delta)$ is an eigenvalue, then also $-E(\lambda,\delta)$ is an eigenvalue. Note that another unitary
transformation $\hat{U}=e^{-i\pi\hat{I}_2}\,e^{-i\pi\hat{J}_1}$ leads to the same conclusion, since the Hamiltonian (\ref{H-lambda-delta-def})
is invariant under an interchange $\hat{\vec{I}} \leftrightarrow \hat{\vec{J}}$.

The matrix elements of $\hat{H}(\lambda,\delta)$ in a given $(I_{\rm T},J_{\rm T})$ sector can be trivially calculated with the aid of
familiar formulas for the involved angular momentum operators. Let us write them down here explicitly for the sake of maximum clarity:
\begin{eqnarray}
%  --------------------------------------------------------------------------------------------------------------------
   \hspace*{-3.00cm} & & \la I_{\rm T}\,I_3\,J_{\rm T}\,J_3 | \, \hat{H}(\lambda,\delta) \,
   | I_{\rm T}\,I'_3\,J_{\rm T}\,J'_3 \ra \; = \nonumber\\
%  --------------------------------------------------------------------------------------------------------------------
   \hspace*{-3.00cm} & = & \delta_{I_3 I'_3} \, \delta_{J_3 J'_3} \; \omega \, \Bigl( I_3 + J_3 \Bigr) \\
%  --------------------------------------------------------------------------------------------------------------------
   \hspace*{-3.00cm} & + & \delta_{I_3 (I'_3+1)} \, \delta_{J_3 (J'_3-1)} \; \lambda \, \phantom{\delta} \,
   \gamma_+\n(I_{\rm T},I'_3) \, \gamma_-\n(J_{\rm T},J'_3) \; + \;
   \delta_{I_3 (I'_3-1)} \, \delta_{J_3 (J'_3+1)} \; \lambda \, \phantom{\delta} \,
   \gamma_-\n(I_{\rm T},I'_3) \, \gamma_+\n(J_{\rm T},J'_3) \nonumber\\
%  --------------------------------------------------------------------------------------------------------------------
   \hspace*{-3.00cm} & + & \delta_{I_3 (I'_3+1)} \, \delta_{J_3 (J'_3+1)} \; \lambda \, \delta \,
   \gamma_+\n(I_{\rm T},I'_3) \, \gamma_+\n(J_{\rm T},J'_3) \; + \;
   \delta_{I_3 (I'_3-1)} \, \delta_{J_3 (J'_3-1)} \; \lambda \, \delta \,
   \gamma_-\n(I_{\rm T},I'_3) \, \gamma_-\n(J_{\rm T},J'_3) \mez ; \nonumber
%  --------------------------------------------------------------------------------------------------------------------
\end{eqnarray}
where by definition
\be
   \gamma_\pm(l,l'\n) \; = \; \sqrt{l(l+1)-l'\n(l'\n\pm1)} \mez .
\ee

In the case of $\delta=0$, the sum $K=(I_3+J_3)$ becomes another good quantum number. Correspondingly,
the $(I_{\rm T},J_{\rm T})$ sector is divided into subsectors associated with $K=(-I_{\rm T}-J_{\rm T}),(-I_{\rm T}-J_{\rm T}+1),\cdots,(+I_{\rm T}+J_{\rm T})$. Moreover, $\hat{H}_0$ commutes both with $\hat{I}_+ \, \hat{J}_-$ and with
$\hat{I}_- \, \hat{J}_+$, hence
\be \label{hat-H-0-hat-V-0-CR}
   \Bigl[ \hat{H}_0 \, , \, \hat{V}_0 \Bigr] \; = \; \hat{0} \mez ;
\ee
exactly as mentioned in the Introduction.
Thereby an eigenvalue problem of the Hamiltonian
\be \label{hat-H-lambda-0}
   \hat{H}(\lambda,0) \; = \; \hat{H}_0 \; + \; \lambda \, \hat{V}_0
\ee
is solvable trivially, provided only that an eigenproblem of $\hat{V}_0$ has been resolved. Accordingly, all the EPs
of $\hat{H}(\lambda,0)$ are trivially known (see the $\delta=0$ panels of Figs.~1 and 2 below, which consist just of intersecting straight lines).
This confirms that our definition of the augmented Hamiltonian (\ref{H-lambda-delta-def}) satisfies the general requirements imposed
on $\hat{H}(\lambda,0)$ in the Introduction and in {\sl Appendix A}.

For illustration, let us present now explicitly the calculated eigenvalue spectrum of $\hat{H}(\lambda,\delta)$
for $N=19$, $\lambda \in [0,1]$, $\delta \in [0,1]$, {\sl even} parity of $(I_3+J_3)$, and $\omega=1.0$. The obtained results are shown in Fig.~1.
An analogous case of {\sl odd} parity is then depicted in Fig.~2.
\begin{figure}[h!]
%\hspace*{+1.00cm}
\includegraphics[angle=0,scale=1.3]{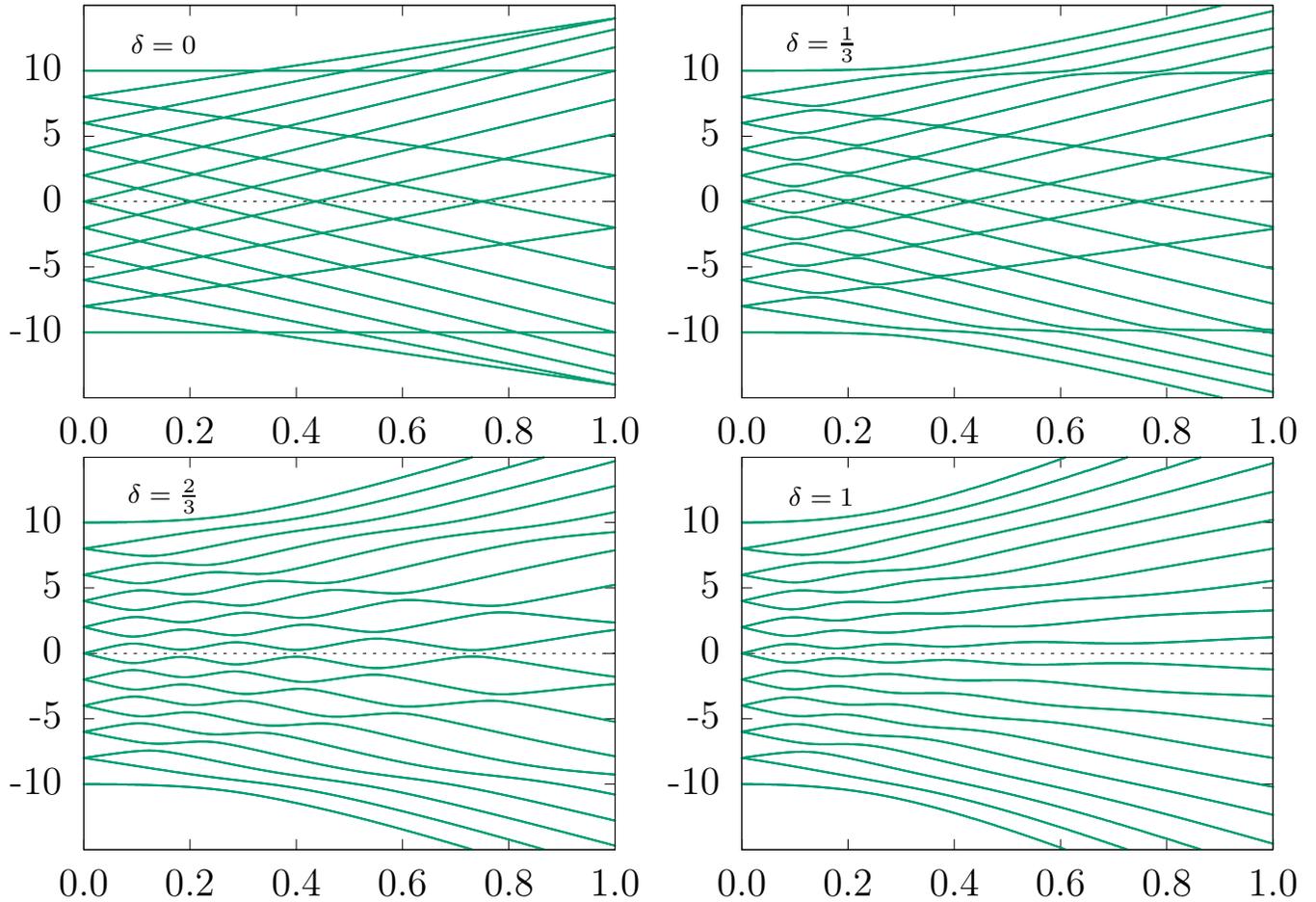}
\caption{ The calculated eigenvalue spectrum of $\hat{H}(\lambda,\delta)$ of Eq.~(\ref{H-lambda-delta-def}) for $N=19$,
{\sl even} parity of $(I_3+J_3)$, and $\omega=1.0$. Horizontal axis corresponds to $\lambda$, vertical axis to the energy variable $E$
associated with the eigenvalues. Note the reflection symmetry of the spectrum with respect to the horizontal $E=0$ axis. This kind of
symmetry is explained by equation (\ref{U-dagger-H-U}) above.}
\end{figure}
\begin{figure}[h!]
%\hspace*{+1.00cm}
\includegraphics[angle=0,scale=1.3]{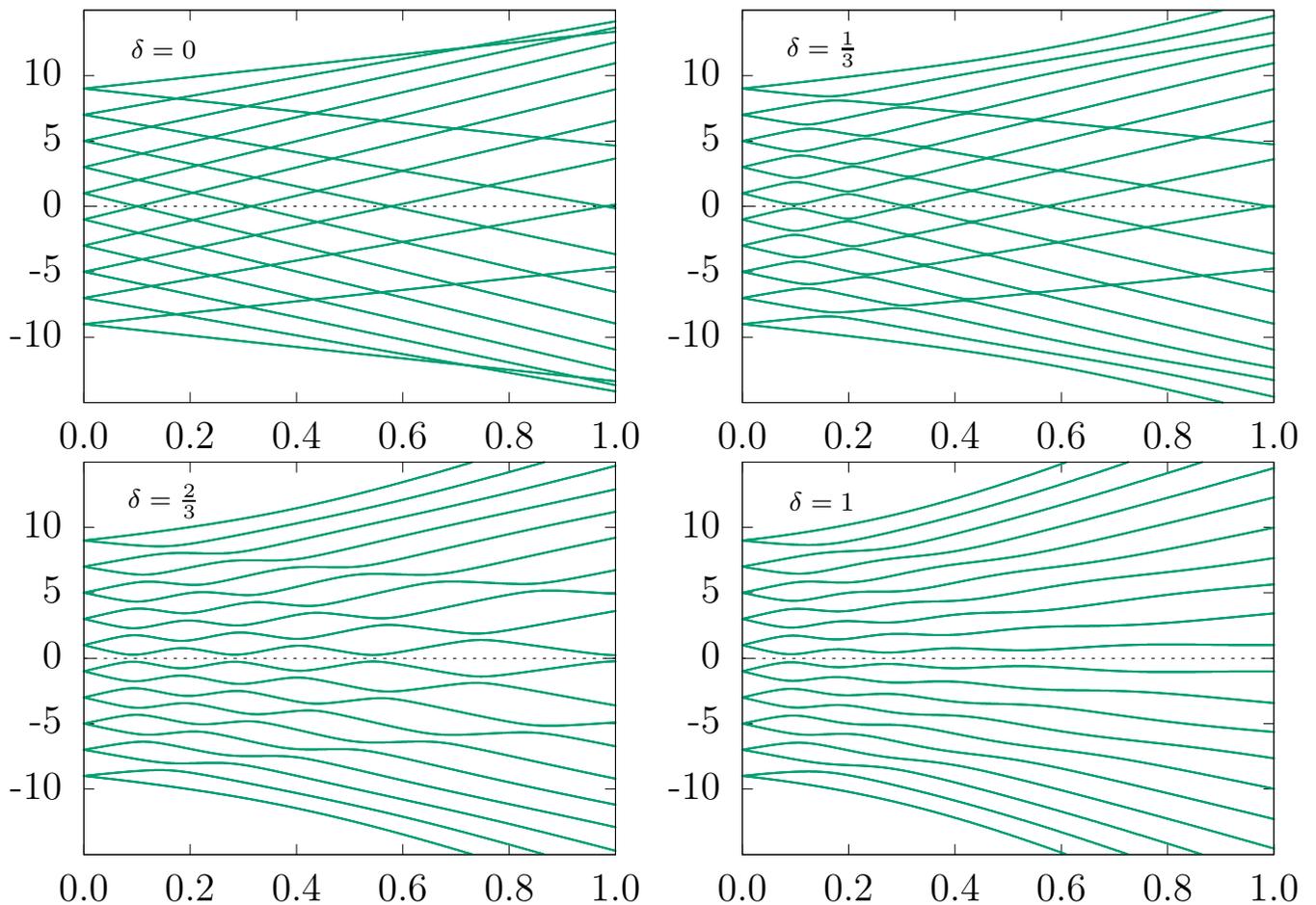}
\caption{ The calculated eigenvalue spectrum of $\hat{H}(\lambda,\delta)$ of Eq.~(\ref{H-lambda-delta-def}) for $N=19$,
{\sl odd} parity of $(I_3+J_3)$, and $\omega=1.0$. Horizontal axis corresponds to $\lambda$, vertical axis to the energy variable $E$
associated with the eigenvalues. Note the reflection symmetry of the spectrum with respect to the horizontal $E=0$ axis. This kind of
symmetry is explained by equation (\ref{U-dagger-H-U}) above.}
\end{figure}

The sought EPs of our starting Hamiltonian $\hat{H}(\lambda)$ of Eq.~(\ref{hat-H-lambda-1-def}) can be identified now with the EPs of
$\hat{H}(\lambda,\delta)$ of Eq.~(\ref{H-lambda-delta-def}) at $\delta=1$. Yet the EPs of
$\hat{H}(\lambda,\delta)$ are obtainable numerically from the hermitian straight line
crossings of $\hat{H}(\lambda,0)$ of Eq.~(\ref{hat-H-lambda-0}) via the parametric $\delta$-propagation
$(\delta=0 \mapsto \delta=1)$ of the EOM, exactly as we formulated in a self contained
fashion in the above Section 2 and in {\sl Appendix A}.

\subsection{Numerical propagation of the EOM and the obtained results}

The seven mutually coupled equations of motion (\ref{dot-lambda-EOM}), (\ref{dot-E-m-final}), (\ref{dot-E-j-final}), (\ref{dot-f-m-EOM}), (\ref{dot-c-m-EOM}), (\ref{dot-c-j-EOM}), (\ref{dot-b-m-EOM}) derived in {\it Subsection 2.2} are propagated numerically using the simplest possible first order difference scheme, starting from the
initial conditions which are established in {\sl Appendix A}. At each propagation step, an internal consistency of the obtained results is strictly checked. Namely, the seven entities (\ref{entities}) calculated for a given
particular value of $\delta \in [0,1]$ are required to satisfy (up to a prescribed numerical accuracy) the three eigenvalue
equations (\ref{H-delta-EVP-M-EPs}), (\ref{H-delta-EVP-ordinary}), (\ref{H-delta-EVP-complementary}), the six orthonormality relations
(\ref{onrel-c-j-c-j'})-(\ref{onrel-b-m-b-m'}), and the closure property (\ref{closure}). In this manner our numerical results presented below
are granted to be reliably converged.

Our illustrative numerical calculations are performed for the toy model of {\it Subsection 3.1}, assuming $N=19$ and $\omega=1$ much as in Figs.~1-2.
The range $[0,1]$ of $\delta$ is discretized by $G=10^7$ equidistant grid points. This ensures that our aforementioned test relations
(\ref{H-delta-EVP-M-EPs}), (\ref{H-delta-EVP-ordinary}), (\ref{H-delta-EVP-complementary}), (\ref{onrel-c-j-c-j'})-(\ref{onrel-b-m-b-m'}),
(\ref{closure}) are fulfilled at each value of $\delta$ with the maximum error not exceeding $0.0005$.

\subsubsection{Results for the odd parity}

\phantom{bla}
\vspace*{+0.20cm}

For maximum clarity of the presentation, it is convenient to start with discussing our results obtained for the case
of odd parity. Our propagation starts from the hermitian crossings which are indicated by red bullets in Fig.~3.
\begin{figure}[h!]
\hspace*{+1.00cm}
\includegraphics[angle=0,scale=1.1]{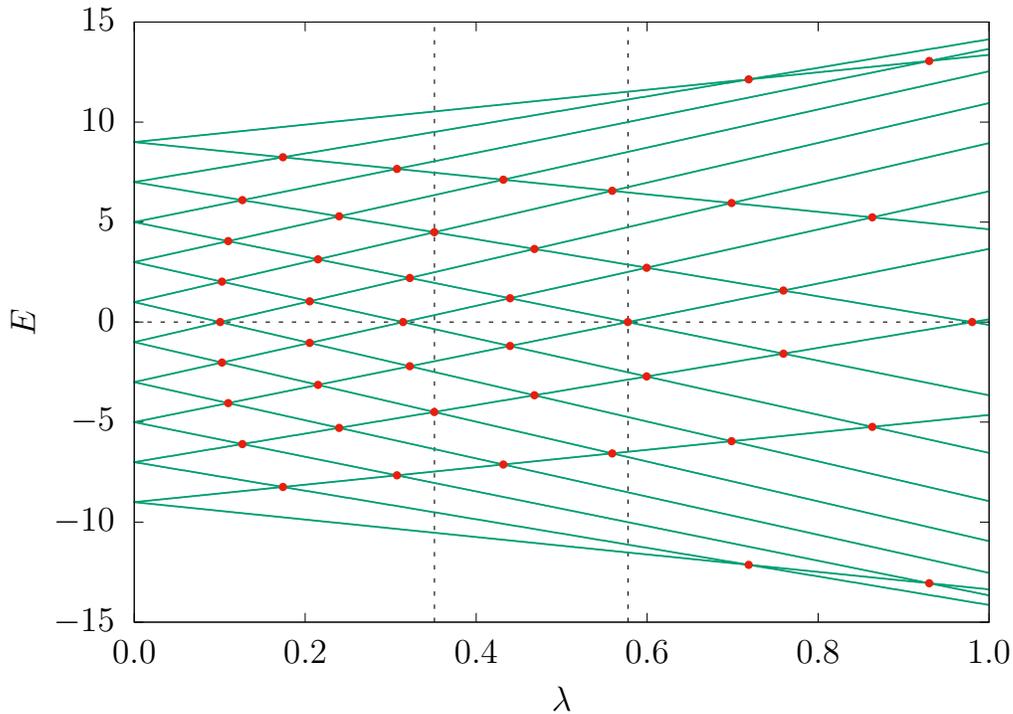}
\caption{ The hermitian straight line crossings corresponding to $\hat{H}(\lambda,0)$, again for $N=19$,
{\sl odd} parity of $(I_3+J_3)$, and $\omega=1.0$. One may observe that there often (through not always)
exist multiple (twofold) crossings for a given value of $\lambda$ (see the vertical black dashed lines).
These are exactly the multiplets described theoretically in {\sl Appendix A}.}
\end{figure}

Our explicit numerical propagation of the EOM provides the following outcomes:
\begin{itemize}
\item[$\star$]
Each isolated (onefold) hermitian crossing of Fig.~3 provides for $\delta>0$ an isolated (onefold) EP.
\item[$\star$]
Each twofold hermitian crossing of Fig.~3 provides for $\delta>0$ the corresponding pair (twofold cluster) of distinct binary EPs
which share the same dependence $\lambda(\delta)$. This is a direct consequence of the symmetry of $\hat{H}(\lambda,\delta)$ which
is highlighted by equation (\ref{U-dagger-H-U}) above.
\end{itemize}
Figs.~4, 5 and 6 present explicitly our most important numerical results, namely, the trajectories of the EPs in the $\lambda$-plane
and in the plane of complex energy. Note that Fig.~4 and Fig.~5 display essentially the same data, just with a different layout convention.
Specifically, Fig.~4 is plotted using a random (machine generated) sign convention for the imaginary part of each obtained curve
$\lambda_k(\delta)$, whereas Fig.~5 corresponds to imposing a fixed convention of $\Im \lambda_k(\delta) \ge 0$.\footnote{
Recall in this context that each curve $\lambda_k(\delta)$ displayed in Figs.~4-5 gives rise to another legitimate curve $\lambda^*(\delta)$
which penetrates into the opposite side of the imaginary $\lambda$-plane, this curve $\lambda_k^*(\delta)$ is not plotted. }
Compared to Fig.~4, the overall appearance of Fig.~5 is somewhat less transparent. In particular, some curves $\lambda_k(\delta)$ do
intersect (albeit at mutually distinct values of $\delta$). This is the only reason why we hereafter prefer to present all our numerical
results using the layout convention analogous to Fig.~4.

\begin{figure}[h!]
\hspace*{+1.00cm}
\includegraphics[angle=0,scale=1.1]{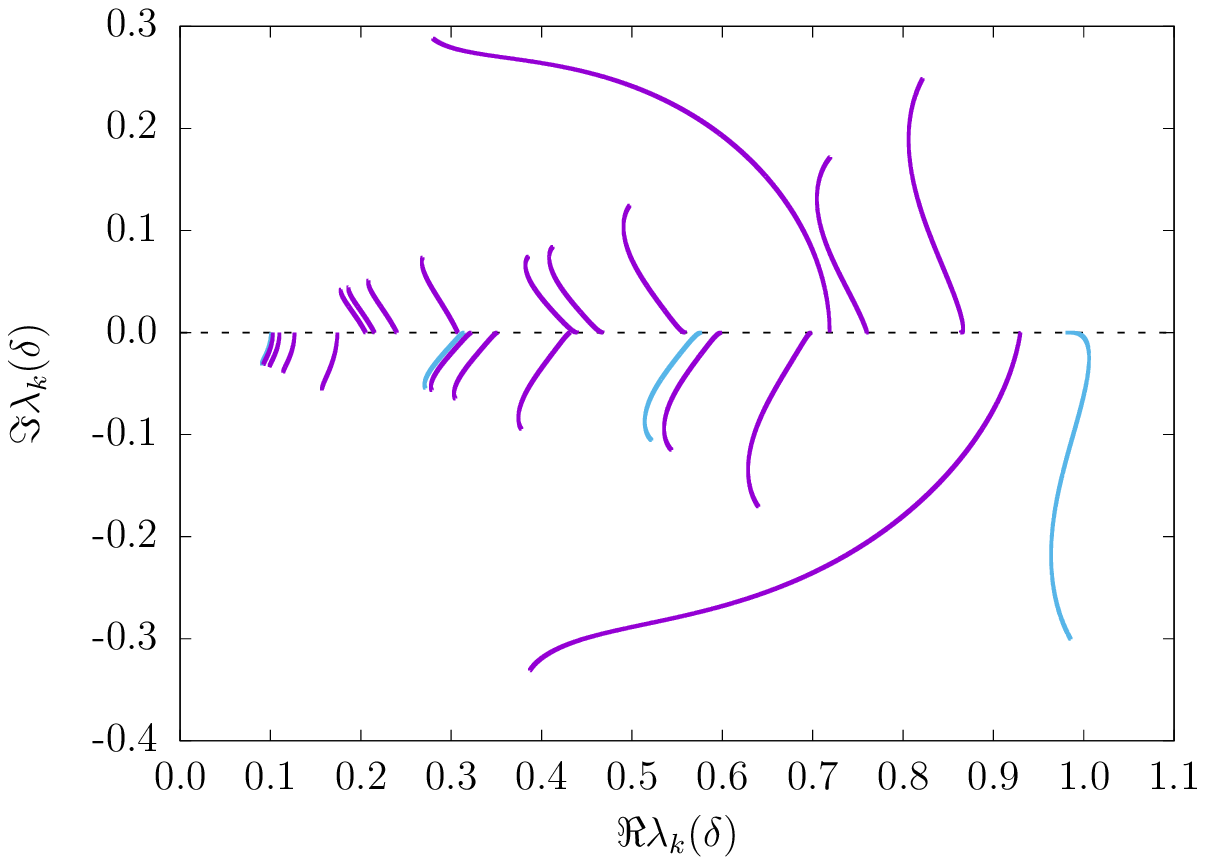}
\caption{ The EP trajectories $\lambda_k(\delta)$ emanating from the hermitian straight line crossings of Fig.~3.
The dark blue trajectories correspond to a pair (twofold cluster) of distinct binary EPs which share the same $\lambda_k(\delta)$,
see our discussion in the main text. On the other hand, the light blue trajectories are associated with a single binary EP.
Note also that each curve $\lambda_k(\delta)$ gives rise to another legitimate curve $\lambda_k^*(\delta)$, which departs from the
same (cluster of) red bullet(s) of Fig.~3, but which corresponds to the complex conjugated initial conditions at $\delta=0$.
$[\,$This means that each curve $\lambda_k(\delta)$ plotted explicitly here in the present figure has been obtained via adopting a particular
(machine generated) sign convention for the $\sigma_1$-factor from {\sl Appendix A}.$\,]$}
\end{figure}

\begin{figure}[h!]
\hspace*{+1.00cm}
\includegraphics[angle=0,scale=1.1]{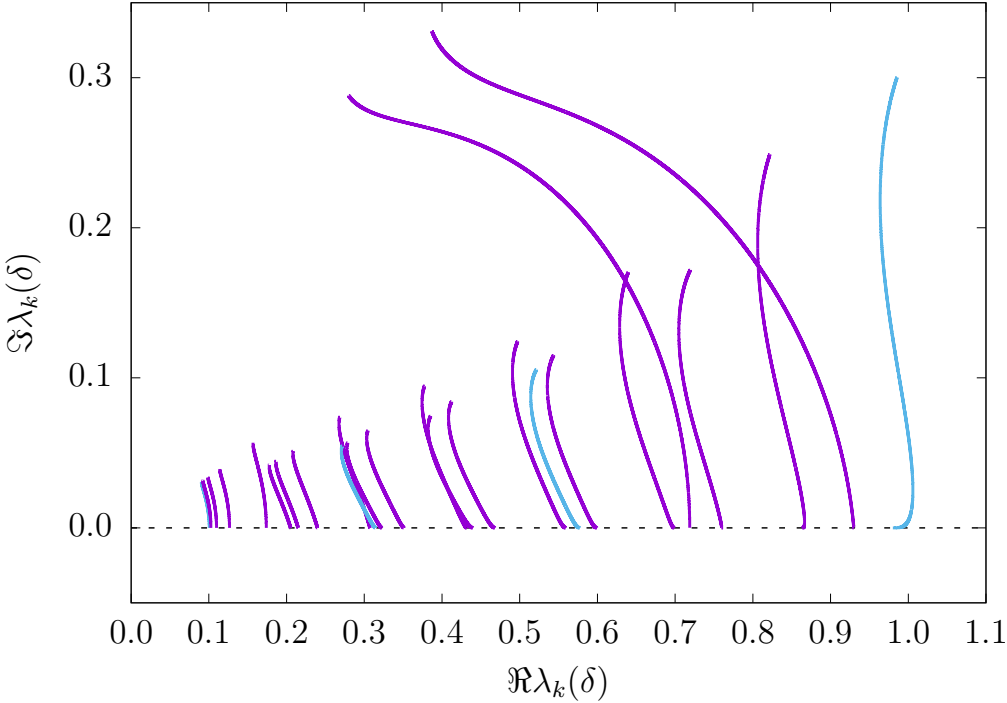}
\caption{ The same data as in Fig.~4, just a sign convention of $\Im \lambda_k(\delta) \ge 0$ is imposed a posteriori.
Note that each curve $\lambda_k(\delta)$ displayed here gives rise to another legitimate curve $\lambda_k^*(\delta)$ which
penetrates into the negative imaginary plane of $\lambda$. Compared to Fig.~4, the overall appearance of the present figure is somewhat less
transparent. In particular, some curves $\lambda_k(\delta)$ do intersect (albeit at mutually distinct values of $\delta$).
This is the only reason why we hereafter prefer to display all our numerical results using the layout convention analogous to Fig.~4. }
\end{figure}

\begin{figure}[h!]
\hspace*{+1.00cm}
\includegraphics[angle=0,scale=1.0]{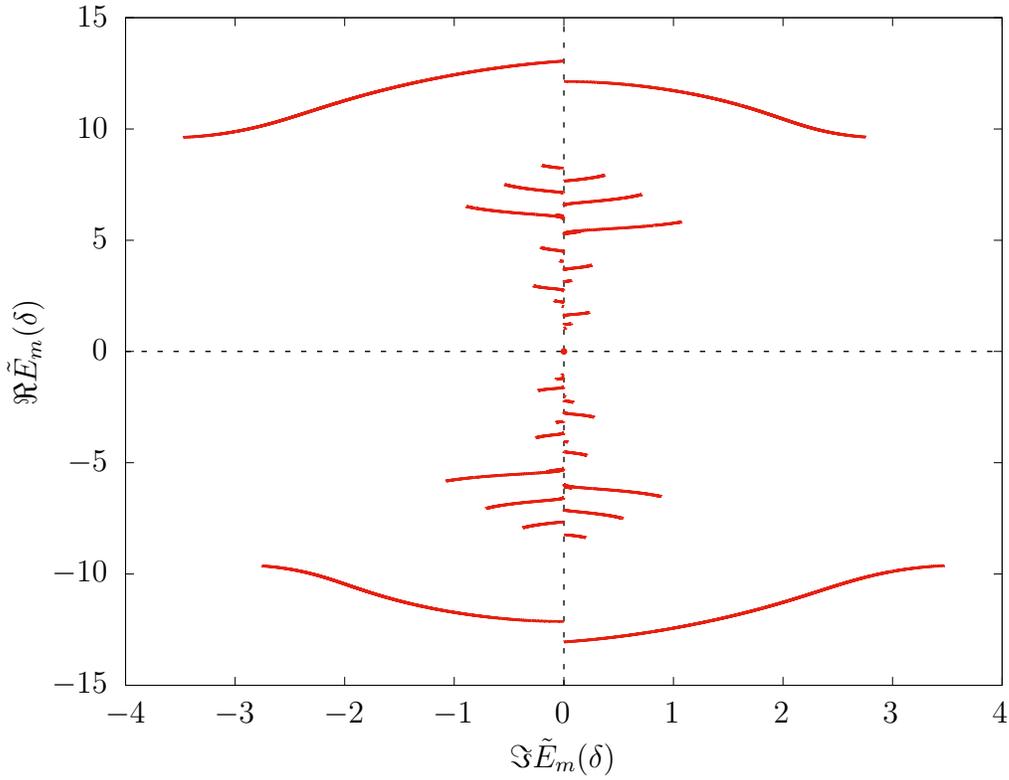}
\caption{ The EP trajectories $\tilde{E}_m(\delta)$ emanating from the hermitian straight line crossings of Fig.~3 and corresponding
          to all the curves $\lambda_k(\delta)$ plotted explicitly in Fig.~4. Importantly, all the onefold hermitian crossings of Fig.~3
          are associated with $\tilde{E}_m(0)=0$, and actually provide $\tilde{E}_m(\delta)=0$ for all $\delta \in [0,1]$. This fact
          (arising as a trivial consequence of the symmetry property (\ref{U-dagger-H-U}) of $\hat{H}(\lambda,\delta)$) is highlighted
          by the presence of red bullet at the origin of the energy plane. On the other hand, the present figure depicts also a progression of several
          nonzero trajectories $\tilde{E}_m(\delta)$, which possess reflection symmetry with respect to the origin.
          Each pair of these symmetry related trajectories corresponds inevitably to a pair (twofold cluster) of distinct binary EPs which
          share the same $\lambda_k(\delta)$. }
\end{figure}

\clearpage

\subsubsection{Results for the even parity}

\phantom{bla}
\vspace*{+0.20cm}

Let us move now on to the case of even parity. Our propagation starts from the hermitian crossings which are indicated by red bullets in Fig.~7.
\begin{figure}[h!]
\hspace*{+1.00cm}
\includegraphics[angle=0,scale=1.1]{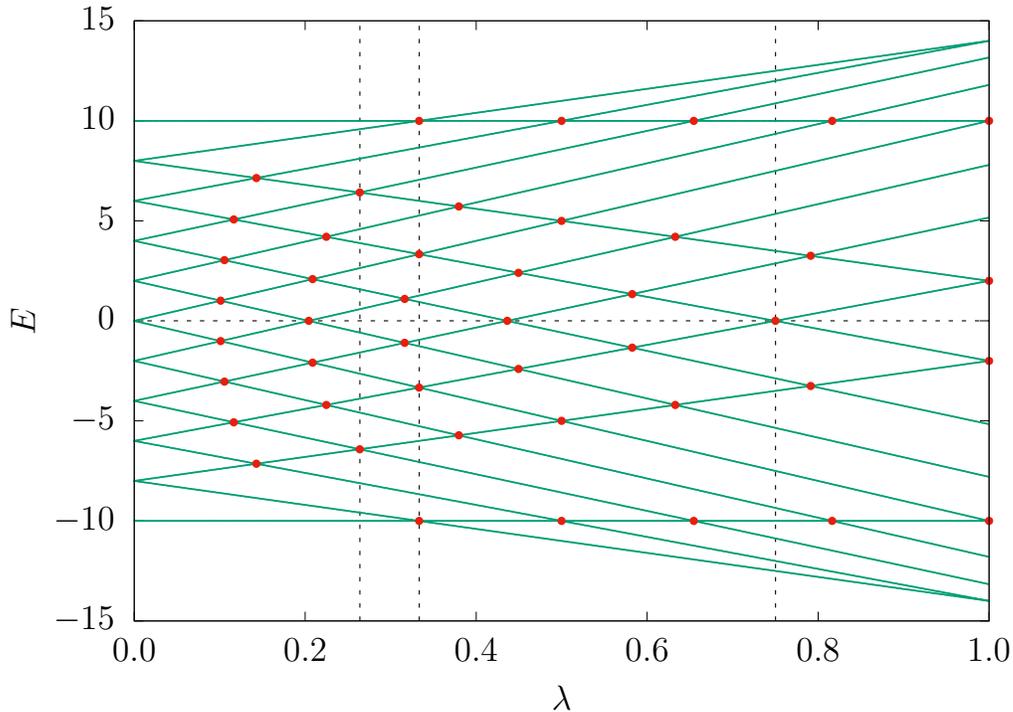}
\caption{ The hermitian straight line crossings corresponding to $\hat{H}(\lambda,0)$, again for $N=19$,
{\sl even} parity of $(I_3+J_3)$, and $\omega=1.0$. One may observe that there often (through not always)
exist multiple (twofold, fourfold) crossings for a given value of $\lambda$ (see the vertical black dashed lines).
These are exactly the multiplets described theoretically in {\sl Appendix A}.}
\end{figure}

Our explicit numerical propagation of the EOM provides the following outcomes:
\begin{itemize}
\item[$\star$]
Each isolated (onefold) hermitian crossing of Fig.~7 provides for $\delta>0$ an isolated (onefold) EP.
\item[$\star$]
Each twofold hermitian crossing of Fig.~7 provides for $\delta>0$ the corresponding pair (twofold cluster) of distinct binary EPs
which share the same dependence $\lambda(\delta)$. This is a direct consequence of the symmetry of $\hat{H}(\lambda,\delta)$ which
is highlighted by equation (\ref{U-dagger-H-U}) above.
\item[$\star$]
The fourfold crossings behave for $\delta>0$ as two separate twofold crossings. Each of these twofold crossings reflects again the
symmetry property (\ref{U-dagger-H-U}) of $\hat{H}(\lambda,\delta)$.
\end{itemize}

Figs.~8 and 9 present explicitly our most important numerical results, namely, the trajectories of the EPs in the $\lambda$-plane
and in the plane of complex energy. We use the same layout convention as in Fig.~4 above.
\begin{figure}[h!]
\hspace*{+1.00cm}
\includegraphics[angle=0,scale=1.1]{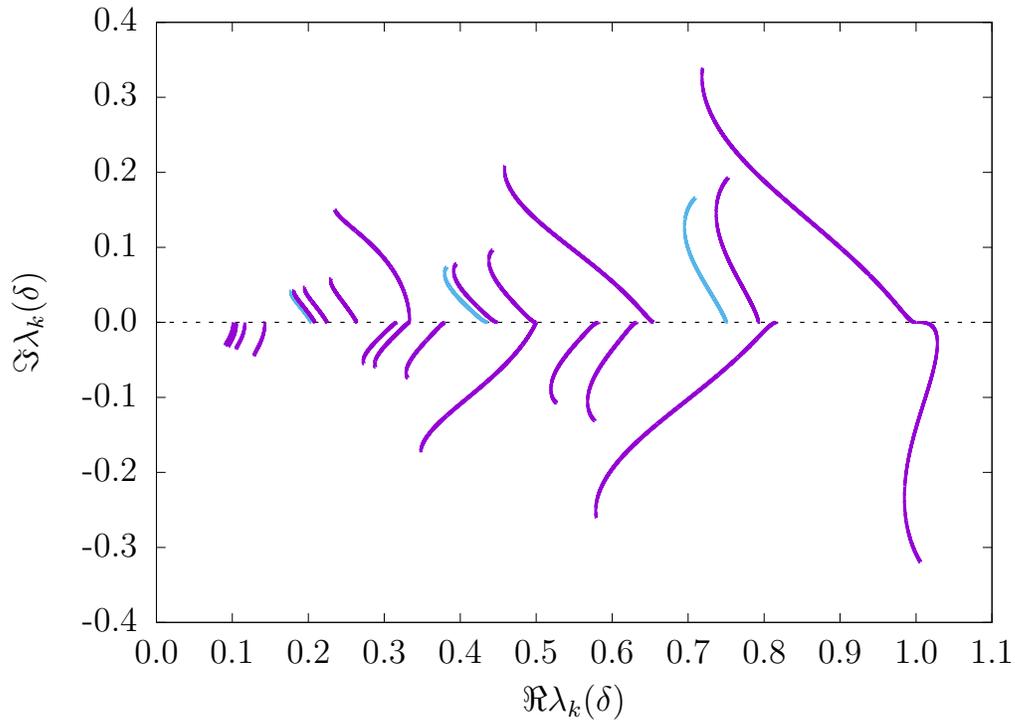}
\caption{
The EP trajectories $\lambda_k(\delta)$ emanating from the hermitian straight line crossings of Fig.~7.
The dark blue trajectories correspond to a pair (twofold cluster) of distinct binary EPs which share the same $\lambda_k(\delta)$,
see our discussion in the main text. On the other hand, the light blue trajectories are associated with a single binary EP.
Note also that each curve $\lambda_k(\delta)$ gives rise to another legitimate curve $\lambda_k^*(\delta)$, which departs from the
same (cluster of) red bullet(s) of Fig.~7, but which corresponds to the complex conjugated initial conditions at $\delta=0$.
$[\,$This means that each curve $\lambda_k(\delta)$ plotted explicitly here in the present figure has been obtained via adopting a particular
(machine generated) sign convention for the $\sigma_1$-factor from {\sl Appendix A}.$\,]$}
\end{figure}
\begin{figure}[h!]
\hspace*{+1.00cm}
\includegraphics[angle=0,scale=1.0]{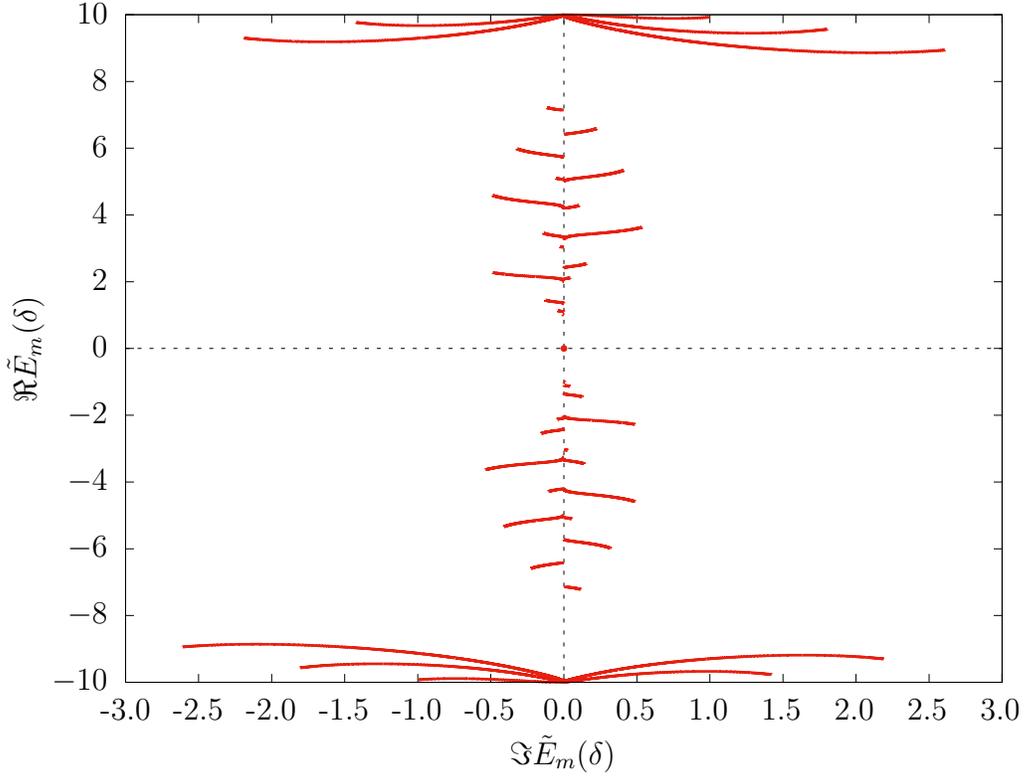}
\caption{ The EP trajectories $\tilde{E}_m(\delta)$ emanating from the hermitian straight line crossings of Fig.~7 and corresponding
          to all the curves $\lambda_k(\delta)$ plotted explicitly in Fig.~8. Importantly, all the onefold hermitian crossings of Fig.~7
          are associated with $\tilde{E}_m(0)=0$, and actually provide $\tilde{E}_m(\delta)=0$ for all $\delta \in [0,1]$. This fact
          (arising as a trivial consequence of the symmetry property (\ref{U-dagger-H-U}) of $\hat{H}(\lambda,\delta)$) is highlighted
          by the presence of red bullet at the origin of the energy plane. On the other hand, the present figure depicts also a progression of several
          nonzero trajectories $\tilde{E}_m(\delta)$, which possess reflection symmetry with respect to the origin.
          Each pair of these symmetry related trajectories corresponds inevitably to a pair (twofold cluster) of distinct binary EPs which
          share the same $\lambda_k(\delta)$. }
\end{figure}

\clearpage

Summarizing the contents of Section 3, we have employed a nontrivial toy model to explicitly test the performance of our
computational algorithm based upon solving the EOM for the EPs. We hope that our illustrative calculations demonstrate practical
usefulness of our EOM method for finding the EPs of nontrivial Hamiltonians.

\section{Concluding remarks}

In summary, the present article establishes the equations of motion (EOM) governing the dynamics (or flux) of EPs of parameterically dependent
nonhermitian Hamiltonians.
This motion of EPs in the parameter space is triggered here by a continuous change of an additional external control parameter of the Hamiltonian. Our analysis covers a
relatively broad class of problems (\ref{hat-H-lambda-take-1}), where the search for EPs can be reinterpreted as solution of EOM pertaining to an
augmented Hamiltonian $\hat{H}\n(\lambda,\delta)$ of Eq.~(\ref{hat-H-lambda-take-3}), with $\delta$ playing the role of the dynamical "time".

From the theoretical point of view, Section 2 represents the most important new hardcore material brought in by our paper.
The resulting EOM (\ref{dot-lambda-EOM}), (\ref{dot-E-m-final}), (\ref{dot-E-j-final}), (\ref{dot-f-m-EOM}), (\ref{dot-c-m-EOM}),
(\ref{dot-c-j-EOM}), (\ref{dot-b-m-EOM}) are based essentially upon implementing a nontraditional perturbation theory of nonhermitian quantum
mechanics in the presence of multiple EPs. An elaboration of such EOM, and in particular derivation of equation (\ref{dot-lambda-EOM}),
brings further theoretical insights into the properties of the EPs, and represents thus a contribution of its own right.

Furthermore, our EOM can be exploited even in a totally pragmatic fashion, merely as an efficient numerical tool for obtaining all the EPs of
interest for a given Hamiltonian $\hat{H}\n(\lambda)$ of Eq.~(\ref{hat-H-lambda-take-1}). Such an approach lends itself for its immediate application e.g.~whenever the sought EPs emanate from avoided crossings of the particular hermitian Hamiltonian
under study. Section 3 demonstrates very explicitly practical merits of our EOM method in the just mentioned situation.

We hope that the EOM formalism developed here can motivate or facilitate further studies of EPs in atomic, nuclear, optical and
condensed matter physics.\\

\vspace*{+0.20cm}

{\bf Acknowledgements}\\

We acknowledge financial support of the Czech Science Foundation under grant Nos.~20-21179S (M.~\v{S}.) and
20-09998S (P.~S.~and P.~C.), and of the Charles University in Prague under project UNCE/SCI/013 (P.~S.~and P.~C.).\\

\clearpage

\hspace*{-0.90cm} {\bf References}\\

%\clearpage

\vspace*{+0.50cm}

\appendix

\section{Initial conditions for the EOM}

Equations of motion (\ref{dot-lambda-EOM}), (\ref{dot-E-m-final}), (\ref{dot-E-j-final}), (\ref{dot-f-m-EOM}), (\ref{dot-c-m-EOM}),
(\ref{dot-c-j-EOM}), (\ref{dot-b-m-EOM}) need to be supplemented with appropriate initial conditions (ICS), i.e., by the seven fundamental
entities (\ref{entities}) provided at some starting value of $\delta=\delta_{\rm in}$. The choice of $\delta_{\rm in}$ is of course governed
by concrete nature of the problem under study. In the present paper, we shall describe a relatively frequently encountered situation when the mentioned
ICS are determinable at $\delta_{\rm in}$ (semi)trivially due to a particularly simple form of $\hat{H}(\lambda,\delta_{\rm in})$. Namely,
we shall be concerned with such an arrangement when the starting Hamiltonian $\hat{H}(\lambda,\delta_{\rm in})$ ($\lambda \in {\mathbb R}$)
of the studied physical model is hermitian (actually, even real symmetric), and possesses exact crossings (accidental degeneracies).
These crossings play the role of origins from which our sought EPs emanate into the complex $\lambda$-plane as $\delta$ is set to depart continuously from $\delta_{\rm in}$.

Let us assume for now $\lambda \in {\mathbb R}$, and consider the hermitian Hamiltonian $\hat{H}(\lambda,\delta_{\rm in})$. Suppose that there
exists some particular value $\lambda_{\rm in} \in {\mathbb R}$ at which the eigenvalue spectrum of $\hat{H}(\lambda,\delta_{\rm in})$ contains
$M_{\rm in}$ simple binary\footnote{ One may of course analyze also more general situations of multiple degeneracies, but this is beyond the
scope of the present paper. } crossings ($1 \leq M_{\rm in} \leq N/2$). Meaning that
\begin{eqnarray} \label{binary-degeneracies}
\;\;\;\;\; E_1\n(\lambda_{\rm in},\delta_{\rm in}) & = & \;\; E_2\n(\lambda_{\rm in},\delta_{\rm in}) \mez ; \nonumber\\
\;\;\;\;\; E_3\n(\lambda_{\rm in},\delta_{\rm in}) & = & \;\; E_4\n(\lambda_{\rm in},\delta_{\rm in}) \mez ; \\
& \;\;\bm\vdots & \nonumber\\
E_{2M_{\rm in}-1}\n(\lambda_{\rm in},\delta_{\rm in}) & = & E_{2M_{\rm in}}\n(\lambda_{\rm in},\delta_{\rm in}) \mz \m ; \nonumber
\end{eqnarray}
where the twice degenerate levels $E_1\n(\lambda_{\rm in},\delta_{\rm in}), \ldots, E_{2M_{\rm in}-1}\n(\lambda_{\rm in},\delta_{\rm in})$
are all distinct, and satisfy also
\begin{eqnarray} \label{binary-slopes}
   \;\;\;\;\; \partial_\lambda \, E_1\n(\lambda,\delta_{\rm in}) \, \Bigr|_{\lambda=\lambda_{\rm in}} & \neq &
   \;\; \partial_\lambda \, E_2\n(\lambda,\delta_{\rm in}) \, \Bigr|_{\lambda=\lambda_{\rm in}} \mez ; \nonumber\\
   \;\;\;\;\; \partial_\lambda \, E_3\n(\lambda,\delta_{\rm in}) \, \Bigr|_{\lambda=\lambda_{\rm in}} & \neq &
   \;\; \partial_\lambda \, E_4\n(\lambda,\delta_{\rm in}) \, \Bigr|_{\lambda=\lambda_{\rm in}} \mez ; \\
   & \;\;\bm\vdots & \nonumber\\
   \partial_\lambda \, E_{2M_{\rm in}-1}\n(\lambda,\delta_{\rm in}) \, \Bigr|_{\lambda=\lambda_{\rm in}} & \neq &
   \partial_\lambda \, E_{2M_{\rm in}}\n(\lambda,\delta_{\rm in}) \, \Bigr|_{\lambda=\lambda_{\rm in}} \mz \m . \nonumber
\end{eqnarray}
Conditions (\ref{binary-slopes}) say that each of the listed "simple binary" crossings corresponds to an intersection of two $\lambda$-dependent
eigenvalue lines with nonequal slopes. All the remaining eigenvalues $E_{j>2M_{\rm in}}\n(\lambda_{\rm in},\delta_{\rm in})$ are assumed
to be nondegenerate. Figs.~1, 2, 3, 7 in the main text illustrate neatly the presence of the just discussed simple binary crossings in the case
of our toy model Hamiltonian at $\delta_{\rm in}=0$. In fact, Figs.~1, 2, 3, 7 depict even several distinct occurrences of $\lambda_{\rm in}$ together with their
pertinent values of $M_{\rm in}$ (one may actually observe that $M_{\rm in} \in \{1,2,4\}$ in these plots).

Let us explore now what happens with a particular multiplet of simple binary crossings $(\delta_{\rm in},\lambda_{\rm in},M_{\rm in})$
once $\delta \in {\mathbb R}$ is set to depart slightly from $\delta_{\rm in}$, and once $\lambda$ is set to deviate
slightly from $\lambda_{\rm in}$ while being allowed to penetrate into the complex plane. As a matter of fact, each crossing
$\kappa \in \{ 1,2,\bm\cdots,M_{\rm in} \}$ survives inside the complex $\lambda$-plane in the form of a binary EP, which moves with
$\delta$ along a certain well defined trajectory $(\delta,\lambda_\kappa(\delta))$. Generally speaking, the resulting trajectories
$\lambda_\kappa(\delta)$ will be $\kappa$-dependent. However, eventual symmetries of $\hat{H}(\lambda,\delta)$ may also cause (some
of) these trajectories to be exactly identical. Under these more peculiar circumstances, our $M_{\rm in}$ binary crossings can be classified
into subgroups (clusters), such that $\lambda_\kappa(\delta)$ is the same within each subgroup (cluster). Different clusters will be hereafter
labeled by index $k$.

Consider now a particular $k$-th cluster of $M$ binary EPs $(1 \leq M \leq M_{\rm in})$. As explained in the previous paragraph,
this cluster of $M$ EPs (whose elements we are going to label by index $m \in \{ 1,2,\bm\cdots,M \}$) emanates from a subset of
simple binary hermitian crossings $(\delta_{\rm in},\lambda_{\rm in},M_{\rm in})$, and $1 \leq M \leq M_{\rm in}$. All the mentioned
$M$ EPs are associated with the same complex $\lambda$-trajectory, $\lambda_k(\delta)$. At $\delta=\delta_{\rm in}$, one has
\be \label{lambda-IC}
   \lambda_k(\delta_{\rm in}) \; = \; \lambda_{\rm in} \mez ;
\ee
and
\begin{eqnarray} \label{binary-degeneracies-EPs}
   \;\;\;\;\; E_1\n(\lambda_{\rm in},\delta_{\rm in}) & = & \;\; E_2\n(\lambda_{\rm in},\delta_{\rm in}) \mez ; \nonumber\\
   \;\;\;\;\; E_3\n(\lambda_{\rm in},\delta_{\rm in}) & = & \;\; E_4\n(\lambda_{\rm in},\delta_{\rm in}) \mez ; \\
   & \;\;\bm\vdots & \nonumber\\
   E_{2M-1}\n(\lambda_{\rm in},\delta_{\rm in}) & = & E_{2M}\n(\lambda_{\rm in},\delta_{\rm in}) \mez \m\m . \nonumber
\end{eqnarray}
We have conveniently adopted here the same kind of notation as above in (\ref{binary-degeneracies}).

Let the orthonormalized eigenvectors corresponding to $E_1\n(\lambda_{\rm in},\delta_{\rm in})$, $E_2\n(\lambda_{\rm in},\delta_{\rm in})$,
$E_3\n(\lambda_{\rm in},\delta_{\rm in})$ etc.~be denoted by symbols $|v_1\n(\lambda_{\rm in},\delta_{\rm in})\ra$,
$|v_2\n(\lambda_{\rm in},\delta_{\rm in})\ra$, $|v_3\n(\lambda_{\rm in},\delta_{\rm in})\ra$, etc. Note that we use here the standard
ket-notation, since $\hat{H}(\lambda_{\rm in},\delta_{\rm in})$ is hermitian (real symmetric) and thus the conventional definition of
the scalar product applies. Since $E_1\n(\lambda_{\rm in},\delta_{\rm in})=E_2\n(\lambda_{\rm in},\delta_{\rm in})$, the sought initial
condition for the $m=1$ EP must inevitably look as follows:
\be \label{IC-E-1-EP}
   \tilde{E}_1^{\delta_{\rm in}} \; = \; E_1\n(\lambda_{\rm in},\delta_{\rm in}) \mez ;
\ee
and
\begin{eqnarray}
%  --------------------------------------------------------------------------------------------------------------------
   \label{IC-c-1-EP}
   | \tilde{c}_1^{\delta_{\rm in}} ) & = & \frac{1}{\sqrt{2}} \, \Bigl( |v_1\n(\lambda_{\rm in},\delta_{\rm in})\ra \, + \,
   \sigma_1 \, i \, |v_2\n(\lambda_{\rm in},\delta_{\rm in})\ra \Bigr) \mez ;\\
%  --------------------------------------------------------------------------------------------------------------------
   \label{IC-b-1-EP}
   | \tilde{b}_1^{\delta_{\rm in}} ) & = & \frac{1}{\sqrt{2}} \, \Bigl( |v_1\n(\lambda_{\rm in},\delta_{\rm in})\ra \, - \,
   \sigma_1 \, i \, |v_2\n(\lambda_{\rm in},\delta_{\rm in})\ra \Bigr) \mez .
%  --------------------------------------------------------------------------------------------------------------------
\end{eqnarray}
In (\ref{IC-c-1-EP})-(\ref{IC-b-1-EP}), the sign factor $\sigma_1 \in \{ -1,+1 \}$. Similarly for all the other EPs $m=2,3,\bm\cdots,M$.
Written down explicitly, we set
\be \label{IC-E-m-EP}
   \tilde{E}_m^{\delta_{\rm in}} \; = \; E_{2m-1}\n(\lambda_{\rm in},\delta_{\rm in}) \mez ;
\ee
and
\begin{eqnarray}
%  --------------------------------------------------------------------------------------------------------------------
   \label{IC-c-m-EP}
   | \tilde{c}_m^{\delta_{\rm in}} ) & = & \frac{1}{\sqrt{2}} \, \Bigl( |v_{2m-1}\n(\lambda_{\rm in},\delta_{\rm in})\ra \, + \,
   \sigma_m \, i \, |v_{2m}\n(\lambda_{\rm in},\delta_{\rm in})\ra \Bigr) \mez ;\\
%  --------------------------------------------------------------------------------------------------------------------
   \label{IC-b-m-EP}
   | \tilde{b}_m^{\delta_{\rm in}} ) & = & \frac{1}{\sqrt{2}} \, \Bigl( |v_{2m-1}\n(\lambda_{\rm in},\delta_{\rm in})\ra \, - \,
   \sigma_m \, i \, |v_{2m}\n(\lambda_{\rm in},\delta_{\rm in})\ra \Bigr) \mez ;
%  --------------------------------------------------------------------------------------------------------------------
\end{eqnarray}
where $1 \leq m \leq M$ and $\sigma_m \in \{ -1,+1 \}$. An assignment of the sign factors $(\sigma_1,\sigma_2,\bm\cdots,\sigma_M)$ in
(\ref{IC-c-m-EP})-(\ref{IC-b-m-EP}) must be performed in such a consistent way that the velocity
\be \label{bm-dot-lambda-consistent}
   \bm\dot{\lambda}(\delta_{\rm in}) \; = \;
   - \, \frac{(\tilde{c}_m^{\delta_{\rm in}}|\partial_\delta\,\hat{H}(\lambda_{\rm in},\delta_{\rm in})|\tilde{c}_m^{\delta_{\rm in}})}
   {(\tilde{c}_m^{\delta_{\rm in}}|\partial_\lambda\,\hat{H}(\lambda_{\rm in},\delta_{\rm in})|\tilde{c}_m^{\delta_{\rm in}})}
\ee
predicted by equation (\ref{dot-lambda-EOM}) comes out as being independent of $m$. We shall return to the sign factors
$(\sigma_1,\sigma_2,\bm\cdots,\sigma_M)$ below (see the item {\it (ii)} in the last paragraph).

We need to specify also the ICS for all the ordinary non-EP eigenstates of $\hat{H}(\lambda_k(\delta_{\rm in}),\delta_{\rm in})$.
This task is straightforward in the case of non-degenerate energy levels $E_{j+2M_{\rm in}}\m(\lambda_{\rm in},\delta_{\rm in})$
(where $1 \leq j \leq N-2M_{\rm in}$). One sets obviously
\be \label{IC-E-j-nondeg}
   E_j^{\delta_{\rm in}} \; = \; E_{j+2M_{\rm in}}\m(\lambda_{\rm in},\delta_{\rm in}) \mez ; \mez 1 \leq j \leq N-2M_{\rm in}
\ee
and
\be \label{IC-c-j-nondeg}
   | c_j^{\delta_{\rm in}} ) \; = \; | v_{j+2M_{\rm in}}\m(\lambda_{\rm in},\delta_{\rm in}) \ra \mez ; \mez 1 \leq j \leq N-2M_{\rm in}
\ee
where $| v_{j+2M_{\rm in}}\m(\lambda_{\rm in},\delta_{\rm in}) \ra$ stands of course for the unit normalized eigenvector of
$\hat{H}(\lambda_{\rm in},\delta_{\rm in})$ associated with level $E_{j+2M_{\rm in}}\m(\lambda_{\rm in},\delta_{\rm in})$.

The situation becomes somewhat more delicate in the case of the doubly degenerate energy eigenvalues listed in (\ref{binary-degeneracies})
but not included in the cluster (\ref{binary-degeneracies-EPs}), namely, in the case of levels
\begin{eqnarray} \label{binary-degeneracies-not-included}
   \;\; E_{2M+1}\n(\lambda_{\rm in},\delta_{\rm in}) & = & E_{2M+2}\n(\lambda_{\rm in},\delta_{\rm in}) \mez ; \nonumber\\
   \;\; E_{2M+3}\n(\lambda_{\rm in},\delta_{\rm in}) & = & E_{2M+4}\n(\lambda_{\rm in},\delta_{\rm in}) \mez ; \nonumber\\
   & \;\;\bm\vdots & \nonumber\\
   E_{2M_{\rm in}-1}\n(\lambda_{\rm in},\delta_{\rm in}) & = & E_{2M_{\rm in}}\n(\lambda_{\rm in},\delta_{\rm in}) \mez \;\,\m . \nonumber
\end{eqnarray}
Consider any given doubly degenerate eigenvalue
\be
   E_{2M+j-N+2M_{\rm in}}\n(\lambda_{\rm in},\delta_{\rm in}) \; = \; E_{2M+j-N+2M_{\rm in}+1}\n(\lambda_{\rm in},\delta_{\rm in})
\ee
where $N-2M_{\rm in}+1 \leq j \leq N-2M-1$. Let the two pertinent unit normalized orthonormal eigenvectors be
\begin{eqnarray}
%  --------------------------------------------------------------------------------------------------------------------------
   \label{v-j-ordinary-deg-listed-1} | v^{(1)} \ra & \equiv &
   | v_{2M+j-N+2M_{\rm in}}\m(\lambda_{\rm in},\delta_{\rm in}) \ra \mez \;\;\;\,\m ;\\
%  --------------------------------------------------------------------------------------------------------------------------
   \label{v-j-ordinary-deg-listed-2} | v^{(2)} \ra & \equiv &
   | v_{2M+j-N+2M_{\rm in}+1}\m(\lambda_{\rm in},\delta_{\rm in}) \ra \mez .
%  --------------------------------------------------------------------------------------------------------------------------
\end{eqnarray}
We set of course
\be \label{IC-E-j-deg}
   E_{2M+j-N+2M_{\rm in}}^{\delta_{\rm in}} \; = \;
   E_{2M+j-N+2M_{\rm in}}\n(\lambda_{\rm in},\delta_{\rm in}) \; = \; E_{2M+j-N+2M_{\rm in}+1}^{\delta_{\rm in}} \mez ;
\ee
much as in (\ref{IC-E-j-nondeg}). Yet an assignment of the corresponding non-EP eigenvectors
\be \label{c-j-ordinary-deg-listed}
   | c_{2M+j-N+2M_{\rm in}}^{\delta_{\rm in}} ) \mez {\rm and} \mez | c_{2M+j-N+2M_{\rm in}+1}^{\delta_{\rm in}} )
\ee
needs a bit more care. Clearly, entities (\ref{c-j-ordinary-deg-listed}) must be built up as $c$-orthonormalized linear combinations
of the two eigenstates (\ref{v-j-ordinary-deg-listed-1})-(\ref{v-j-ordinary-deg-listed-2}). In addition, however, one must ensure that
the two sought non-EP eigenvectors (\ref{c-j-ordinary-deg-listed}) are not mutually coupled by the Hamiltonian $\delta$-derivative
(\ref{V-delta-def}), i.e., by the operator
\be
   \hat{V}\m(\delta_{\rm in}) \; = \;
   \partial_\lambda\,\hat{H}(\lambda_{\rm in},\delta_{\rm in}) \, \bm\dot{\lambda}(\delta_{\rm in})
   \; + \; \partial_\delta\,\hat{H}(\lambda_{\rm in},\delta_{\rm in}) \mez .
\ee
Indeed, the just imposed extra requirement of
\be
   ( c_{2M+j-N+2M_{\rm in}}^{\delta_{\rm in}} | \hat{V}\m(\delta_{\rm in}) | c_{2M+j-N+2M_{\rm in}+1}^{\delta_{\rm in}} ) \; = \; 0
\ee
is indispensable, since it guarantees that our EOM (\ref{dot-c-j-EOM}) does not possess a singularity at $\delta=\delta_{\rm in}$.
Hence an appropriate kind of regularization or rectification must be implemented here. In fact, an explicit construction of the two non-EP
eigenvectors (\ref{c-j-ordinary-deg-listed}) is conceptually straightforward. Namely, we diagonalize\footnote{
The matrix (\ref{V-matrix-2-by-2}) is surely diagonalizable. Since if it was non-diagonalizable, then the just investigated crossing
of eigenvalues $E_{2M+j-N+2M_{\rm in}}\n(\lambda_{\rm in},\delta_{\rm in})=E_{2M+j-N+2M_{\rm in}+1}\n(\lambda_{\rm in},\delta_{\rm in})$
would bring an additional $(M+1)$-th EP into the list (\ref{binary-degeneracies-EPs}), contrary to our starting assumption. }
the 2-by-2 matrix
\be \label{V-matrix-2-by-2}
   \left( \matrix{ \la v^{(1)} | \hat{V}\m(\delta_{\rm in}) | v^{(1)} \ra & \la v^{(1)} | \hat{V}\m(\delta_{\rm in}) | v^{(2)} \ra \cr
   \la v^{(2)} | \hat{V}\m(\delta_{\rm in}) | v^{(1)} \ra & \la v^{(2)} | \hat{V}\m(\delta_{\rm in}) | v^{(2)} \ra } \right) \mez ;
\ee
and access in this way the two associated eigenvectors
\be
   \vec{w}^{(1)} \; = \;
   \left( \matrix{ w_1^{(1)} \cr w_2^{(1)} } \right) \mez , \mez \vec{w}^{(2)} \; = \; \left( \matrix{ w_1^{(2)} \cr w_2^{(2)} } \right) \mez .
\ee
Subsequently we set
\begin{eqnarray}
\label{IC-c-j-deg-1}
| c_{2M+j-N+2M_{\rm in}\phantom{+1}}^{\delta_{\rm in}} ) & = & w_1^{(1)} \, | v^{(1)} \ra \; + \; w_2^{(1)} \, | v^{(2)} \ra \mez ;\\
\label{IC-c-j-deg-2}
| c_{2M+j-N+2M_{\rm in}+1}^{\delta_{\rm in}} ) & = & w_1^{(2)} \, | v^{(1)} \ra \; + \; w_2^{(2)} \, | v^{(2)} \ra \mez ;
\end{eqnarray}
while tacitly implementing the $c$-normalization.

What remains to be done is to supply the values of $f_m^{\delta_{\rm in}}$. Equation (\ref{H-delta-EVP-complementary}) implies
\be \label{ICS-f-m}
   f_m^{\delta_{\rm in}} \; = \; 0 \mez ; \mez 1 \leq m \leq M
\ee
valid simply because $| \tilde{b}_m^{\delta_{\rm in}} )$ is an eigenvector of $\hat{H}(\lambda_k(\delta_{\rm in}),\delta_{\rm in})$
with an eigenvalue $\tilde{E}_m^{\delta_{\rm in}}$\m.

Summarizing, in this {\sl Appendix A} we have described in a self contained fashion how to specify adequately the ICS for the seven
fundamental entities (\ref{entities}). The resulting ICS are given above in equations (\ref{lambda-IC}), (\ref{IC-E-m-EP}), (\ref{IC-c-m-EP}),
(\ref{IC-b-m-EP}), (\ref{IC-E-j-nondeg}), (\ref{IC-c-j-nondeg}), (\ref{IC-E-j-deg}), (\ref{IC-c-j-deg-1}), (\ref{IC-c-j-deg-2}), and
(\ref{ICS-f-m}). The just mentioned ICS must satisfy by construction the basic eigenvalue and eigenvector properties
(\ref{H-delta-EVP-M-EPs}), (\ref{H-delta-EVP-ordinary}), (\ref{H-delta-EVP-complementary}), (\ref{onrel-c-j-c-j'})-(\ref{onrel-b-m-b-m'}),
(\ref{closure}) listed in Section 2 of the main text, this may serve as an useful consistency check.

Let us finally mention two important questions which still need to be addressed:
\vspace*{-0.20cm}
\begin{itemize}
\item[{\it (i)}] A nontrivial puzzle arises on how a given multiplet of simple binary crossings
                 $(\delta_{\rm in},\lambda_{\rm in},M_{\rm in} \ge 2)$ should be split into specific subgroups (clusters) characterized
                 by the same $\lambda_k(\delta)$.
\item[{\it (ii)}] Another nontrivial puzzle concerns consistent choice of the $M$ sign factors $(\sigma_1,\sigma_2,\bm\cdots,\sigma_M)$
                  for a given $k$-th cluster in equations (\ref{IC-c-m-EP})-(\ref{IC-b-m-EP}), see the above discussion of requirement
                  (\ref{bm-dot-lambda-consistent}).
\end{itemize}
\vspace*{-0.20cm}
Both puzzles {\it (i)-(ii)} are resolved correctly iff an explicit solution of our EOM (\ref{dot-lambda-EOM}), (\ref{dot-E-m-final}),
(\ref{dot-E-j-final}), (\ref{dot-f-m-EOM}), (\ref{dot-c-m-EOM}), (\ref{dot-c-j-EOM}), (\ref{dot-b-m-EOM}), starting from the just
discussed ICS, provides unique outcomes (\ref{entities}) which do possess the basic properties (\ref{H-delta-EVP-M-EPs}),
(\ref{H-delta-EVP-ordinary}), (\ref{H-delta-EVP-complementary}), (\ref{onrel-c-j-c-j'})-(\ref{onrel-b-m-b-m'}),
(\ref{closure}) of Section 2 for all values of $\delta$ considered in the propagation. On the other hand, any inconsistency detected
during the propagation of our EOM, manifested e.g.~by violation of any from the properties (\ref{H-delta-EVP-M-EPs}),
(\ref{H-delta-EVP-ordinary}), (\ref{H-delta-EVP-complementary}), (\ref{onrel-c-j-c-j'})-(\ref{onrel-b-m-b-m'}), (\ref{closure}),
would inevitably imply an incorrect resolution of one or both of the aforementioned issues {\it (i)-(ii)}. Hence the two puzzles
{\it (i)-(ii)} can be uniquely resolved simply on the trial-and-error basis, even in such situations when direct answer to {\it (i)-(ii)}
is not a priori obvious.

\end{document}